
%

\def\boringfonts{y}   
\def\figflag{y} \input epsf  
\long\def\mm#1{#1}
\long\def\kk#1{#1}
\long\def\rr#1{#1}

\input harvmac  


\def\fonttest{y}
\font\cmbsy=cmbsy10
\ifx\boringfonts\fonttest\def\ninepoint{}
\else
\font\ninerm=cmr9\font\ninei=cmmi9\font\nineit=cmti9\font\ninesy=cmsy9
\font\ninebf=cmbx9\font\ninesl=cmsl9\font\ninett=cmtt9
\def\ninepoint{\def\rm{\fam0\ninerm}
\textfont0=\ninerm \scriptfont0=\sevenrm \scriptscriptfont0=\fiverm
\textfont1=\ninei  \scriptfont1=\seveni  \scriptscriptfont1=\fivei
\textfont2=\ninesy \scriptfont2=\sevensy \scriptscriptfont2=\fivesy
\textfont\itfam=\nineit \def\it{\fam\itfam\nineit} \def\sl{\fam\slfam\ninesl}
\textfont\bffam=\ninebf \def\bf{\fam\bffam\ninebf}
\def\tt{\fam\ttfam\ninett}\rm}
\fi

\hyphenation{anom-aly anom-alies coun-ter-term coun-ter-terms
dif-feo-mor-phism dif-fer-en-tial super-dif-fer-en-tial dif-fer-en-tials
super-dif-fer-en-tials reparam-etrize param-etrize reparam-etriza-tion}


%
%
%
\newwrite\tocfile\global\newcount\tocno\global\tocno=1
\ifx\bigans\answ \def\tocline#1{\hbox to 320pt{\hbox to 45pt{}#1}}
\else\def\tocline#1{\line{#1}}\fi
\def\toclead{\leaders\hbox to 1em{\hss.\hss}\hfill}
\def\tnewsec#1#2{\newsec{#2}\xdef #1{\the\secno}
\ifnum\tocno=1\immediate\openout\tocfile=toc.tmp\fi\global\advance\tocno
by1
{\let\the=0\edef\next{\write\tocfile{\medskip\tocline{\secsym\ #2\toclead\the
\count0}\smallskip}}\next}
}
\def\tnewsubsec#1#2{\subsec{#2}\xdef #1{\the\secno.\the\subsecno}
\ifnum\tocno=1\immediate\openout\tocfile=toc.tmp\fi\global\advance\tocno
by1
{\let\the=0\edef\next{\write\tocfile{\tocline{ \ \secsym\the\subsecno\
#2\toclead\the\count0}}}\next}
}
\def\tappendix#1#2#3{\xdef #1{#2.}\appendix{#2}{#3}
\ifnum\tocno=1\immediate\openout\tocfile=toc.tmp\fi\global\advance\tocno
by1
{\let\the=0\edef\next{\write\tocfile{\tocline{ \ #2.
#3\toclead\the\count0}}}\next}
}
%
%

%
%
%
%

\gdef\prlmode{F}
\long\def\optional#1{}
\def\cmp#1{#1}         
%
%
\let\narrowequiv=\equiv
\def\equiv{\;\narrowequiv\;}
\let\narrowtilde=\tilde
\def\tilde{\widetilde}
\fontdimen16\tensy=2.7pt\fontdimen17\tensy=2.7pt 



%

\def\la{\lambda}
\def\ep{\epsilon}

\def\dl{\delta}
%
%

\def\CQ{{\cal Q}}

\def\CO{{\cal O}}

\def\CE{{\cal E}}

%
%
%
\def\boxit#1#2{
        $$\vcenter{\vbox{\hrule\hbox{\vrule\kern3pt\vbox{\kern3pt
        \hbox to #1truein{\hsize=#1truein\vbox{#2}}\kern3pt}\kern3pt\vrule}
        \hrule}}$$
}



\font\cmbsy=cmbsy10
\def\bnabla{\hbox{$\textfont2=\cmbsy\mathchar"272$}}  

%

\def\lfr#1#2{{\textstyle{#1\over#2}}} 



\def\splitexact#1#2{\mathrel{\mathop{\null{
\lower4pt\hbox{$\rightarrow$}\atop\raise4pt\hbox{$\leftarrow$}}}\limits
^{#1}_{#2}}}

%
%
\def\pa{\partial}

\def\pd#1#2{{\partial #1\over\partial #2}} 
%
%
%
%
\def\rmi{{\rm i}}

\def\Im{{\rm Im}\,}\def\Re{{\rm Re}\,}  
\def\ex#1{{\rm e}^{#1}}                 
\def\dd{\mskip 1.3mu{\rm d}\mskip .7mu} 



%
%

\def\IM{isomorphism}

\def\eg{{\it e.g.}}\def\ie{{\it i.e.}}\def\via{{\it via}}

%
%

\ifx\boringfonts\fonttest
\font\blackboard=cmssbx10 \font\blackboards=cmssbx10 at 7pt  
\font\blackboardss=cmssbx10 at 5pt
\else
\font\blackboard=msym10 \font\blackboards=msym7   
\font\blackboardss=msym5
\fi
\newfam\black
\textfont\black=\blackboard
\scriptfont\black=\blackboards
\scriptscriptfont\black=\blackboardss


%
\ifx\boringfonts\fonttest
\font\gothic=cmssbx10 \font\gothics=cmssbx10 at 7pt  
\font\gothicss=cmssbx10 at 5pt
\else
\font\gothic=eufm10 \font\gothics=eufm7
\font\gothicss=eufm5
\fi
\newfam\gothi
\textfont\gothi=\gothic
\scriptfont\gothi=\gothics
\scriptscriptfont\gothi=\gothicss

{\catcode`\@=11\gdef\oldcal{\fam\tw@}}
\newfam\curly
\ifx\boringfonts\fonttest\else
\font\curlyfont=eusm10 \font\curlyfonts=eusm7
\font\curlyfontss=eusm5
\textfont\curly=\curlyfont
\scriptfont\curly=\curlyfonts
\scriptscriptfont\curly=\curlyfontss
\def\cal{\fam\curly\relax}
\fi
%

\ifx\boringfonts\fonttest\def\df{\bf}\else\font\df=cmssbx10\fi

\global\newcount\pnfigno \global\pnfigno=1
\newwrite\ffile
\def\pfig#1#2{Fig.~\the\pnfigno\pnfig#1{#2}}
\def\pnfig#1#2{\xdef#1{Fig. \the\pnfigno}%
\ifnum\pnfigno=1\immediate\openout\ffile=figs.tmp\fi%
\immediate\write\ffile{\noexpand\item{\noexpand#1\ }#2}%
\global\advance\pnfigno by1}

%
%
\def\tfig#1{Fig.~\the\pnfigno\xdef#1{Fig.~\the\pnfigno}\global\advance\pnfigno
by1}

%
%
%
%
\def\figI{y}
\def\ifigure#1#2#3#4{
\midinsert
\ifx\figflag\figI
 \ifx\htflag\figI
 \vbox{
  \href{file:#3}
{Click here for enlarged figure.}}
 \fi
 \vbox to #4truein{
 \vfil\centerline{\epsfysize=#4truein\epsfbox{#3}}}
\else
\vbox to .2truein{}
\fi
\narrower\narrower\noindent{\bf #1:} #2
\endinsert
}








%
%

%


\def\inbar{\,\vrule height1.5ex width.4pt depth0pt}
\def\IB{\relax{\rm I\kern-.18em B}}
\def\IC{\relax\hbox{$\inbar\kern-.3em{\rm C}$}}
\def\ID{\relax{\rm I\kern-.18em D}}
\def\IE{\relax{\rm I\kern-.18em E}}
\def\IF{\relax{\rm I\kern-.18em F}}
\def\IG{\relax\hbox{$\inbar\kern-.3em{\rm G}$}}
\def\IH{\relax{\rm I\kern-.18em H}}
\def\II{\relax{\rm I\kern-.18em I}}
\def\IK{\relax{\rm I\kern-.18em K}}
\def\IL{\relax{\rm I\kern-.18em L}}
\def\IM{\relax{\rm I\kern-.18em M}}
\def\IN{\relax{\rm I\kern-.18em N}}
\def\IO{\relax\hbox{$\inbar\kern-.3em{\rm O}$}}
\def\IP{\relax{\rm I\kern-.18em P}}
\def\IQ{\relax\hbox{$\inbar\kern-.3em{\rm Q}$}}
\def\IR{\relax{\rm I\kern-.18em R}}
\font\cmss=cmss10 \font\cmsss=cmss10 at 10truept
\def\IZ{\relax\ifmmode\mathchoice
{\hbox{\cmss Z\kern-.4em Z}}{\hbox{\cmss Z\kern-.4em Z}}
{\lower.9pt\hbox{\cmsss Z\kern-.36em Z}}
{\lower1.2pt\hbox{\cmsss Z\kern-.36em Z}}\else{\cmss Z\kern-.4em Z}\fi}
\def\IGa{\relax\hbox{${\rm I}\kern-.18em\Gamma$}}
\def\IPi{\relax\hbox{${\rm I}\kern-.18em\Pi$}}
\def\ITh{\relax\hbox{$\inbar\kern-.3em\Theta$}}
\def\IOm{\relax\hbox{$\inbar\kern-3.00pt\Omega$}}

\def\micron{$\mu$m}


\long\def\optional#1{}
\def\pagin#1{}

\def\cmp#1{#1 }         

\def\pndate{9/95}
\long\def\suppress#1{}
\suppress{\def\boringfonts{y}  
\def\pndate{\vbox{\vskip .2truein\hbox{{\sl Running title:}
Pearling Instability}\hbox{{\sl PACS: 02.40.-k, 
47.20.-k, 
47.20.Dr, 
47.20.Gv, 
68.10.-m, 
87.22.Bt.} }
}}
\baselineskip=20pt
\def\ifigure#1#2#3#4{\nfig\dumfig{#2}}

} 


\def\testp{T}

\ifx\answ\bigans

\line{\hfil UPR--669}
\rlap{\phantom{.}}\vskip -3.0cm

\Title{\vbox{\hbox{}}}
{\vbox{\centerline{Front Propagation in the Pearling}
\vskip2pt\centerline{Instability of Tubular Vesicles}
}}
\rlap{\phantom{.}}\vskip -1.5cm

\else
\Title{\vbox{\hbox{UPR--669}}}{\vbox{\centerline{Front Propagation in
the Pearling 
Instability of Tubular Vesicles}
}}
\fi
{\centerline{Raymond E. Goldstein}\smallskip
\centerline{Joseph Henry Laboratory of Physics, Princeton University}
\centerline{Princeton, NJ 08544 USA}\bigskip
\centerline{Philip Nelson and Thomas Powers%
\footnote*{Present address: Dept. of Physics,
Princeton University, Princeton, NJ 08544 and
NEC Research Institute, 4 Independence Way, Princeton, NJ 08540.
}}\smallskip
\centerline{Physics Department, University of Pennsylvania}
\centerline{Philadelphia, PA 19104 USA}
\bigskip\centerline{Udo Seifert}\smallskip
\centerline{Max-Planck-Institut f\"ur Kolloid- und Grenzfl\"achenforschung}
\centerline{Kantstr. 55, 14513 Teltow-Seehof, Germany}
}
\bigskip
{\ninepoint
\baselineskip=10pt
Recently Bar-Ziv and Moses discovered a dynamical shape transformation
induced in cylindrical lipid bilayer vesicles by the action of laser
tweezers. We develop a hydrodynamic theory of fluid bilayers in
interaction with the surrounding water and argue that the effect of
the laser is to induce a sudden tension in the membrane. We refine our
previous analysis to account 
for the fact that the shape transformation is not uniform but propagates outward
from the laser trap. Applying the marginal stability criterion to this
situation gives us an improved prediction for the selected initial wavelength
and a new prediction for the propagation velocity, both in rough
agreement with the experimental values. For example, a tubule of
initial radius 0.7\micron\ has a predicted initial sinusoidal
perturbation in its diameter with wavelength 5.5\micron, as observed.
The perturbation propagates as a front with the qualitatively correct
front velocity a bit less than 100\micron/sec. In particular we show why this
velocity is initially constant, as observed\kk{, and so much smaller
than the natural scale set by the tension}. We also predict that the front
velocity should increase linearly with laser power. Finally we
introduce an approximate hydrodynamic model applicable to the fully
nonlinear regime. This model exhibits
propagating fronts as well as fully-developed ``pearled" vesicles similar to
those seen in the experiments.
} 

\ifx\prlmode\testp
\noindent {\sl PACS: 02.40.-k, 
47.20.-k, 
47.20.Dr, 
47.20.Gv, 
68.10.-m, 
87.22.Bt. 
}\fi
\ifx\answ\bigans \else\noblackbox\fi
\Date{\pndate}\noblackbox


\def\hnabla{\hat\nabla}
\def\lap{_{\rm Laplace}}
\def\micron{$\mu$m}

\def\fpd#1#2{{\delta #1\over\delta #2}} 
\def\pd#1#2{{\pa#1\over\pa#2}}
\def\max{_{\rm max}} \def\crit{_{\rm crit}} 
\def\erg{{\rm erg\thinspace}}
\def\const{{\rm const.}}
 
\def\eff{_{\rm eff}}

\def\cmmt{cm$^{-2}$}

\def\ee#1{\cdot 10^{#1}}
\def\eem#1{\cdot 10^{-#1}}
\def\dvt{\narrowtilde v^+-\narrowtilde v^-}
\def\vt{\narrowtilde v}\def\vtz{\narrowtilde v_z}\def\vf{v_f} \def\vz{v_z}
\def\nz{\nabla_z}

\def\Fphys{3}
\def\FFineq{\Fphys.1}
\def\FFlaser{\Fphys.2}

\def\Fvsm{4}
\def\Ffront{5}
\def\Fexact{6}

\def\FFcyl{\Fexact.3}
\def\Fnon{7}

\def\FAmodulus{C}
\def\FAlambda{D}
\def\FAade{E}
\def\elasticsect{Sect.~\FFcyl}
\def\fekinfac{\FAlambda.9}
\def\fecons{\Fvsm.2}              
\def\fenFB{\FAlambda.7}   
\def\fetotstrs{\FAlambda.8} %

\hfuzz=3truept

\lref\rXMSC{G. Ahlers and D.S. Cannell, ``Vortex-front propagation in
rotating Couette-Taylor flow,''Phys. Rev. Lett. {\bf50} (1983)1583\pagin{--1586};
J. Fineberg and V. Steinberg, ``Vortex-front propagation in
Rayleigh-B\'enard flow,'' Phys. Rev. Lett. {\bf58} (1987) 1332\pagin{--1335}.} 
\lref\GrOl{R. Granek and Z. Olami, ``Dynamics of Rayleigh-like
instability induced by laser tweezers in tubular vesicles of
self-assembled membranes,'' J. Phys. II France {\bf5} (1995) 1348\pagin{--1370}.}
\lref\rEvNe{E. Evans and D. Needham, ``Physical properties of
surfactant bilayer membranes: thermal 
transitions, elasticity, rigidity, cohesion, and colloidal
interactions,'' J. Phys. Chem. {\bf91} (1987) 4219.}%
\lref\ShiBrenNagl {X.D. Shi, M.P. Brenner, and S.R. Nagel, 
``A cascade of structure in a drop falling from a faucet,'' Science
{\bf 265}, 219 (1994). }
\lref\rMiHo{H. Miyata and H. Hotani,
``Morphological changes in liposomes caused by polymerization of 
encapsulated actin and spontaneous formation of actin bundles,"
Proc. Nat. Acad. Sci. USA  {\bf89} (1992) 11547\pagin{--11551}.}
\lref\rpinching{R.E. Goldstein, A.I. Pesci, and M.J. Shelley, ``Topology
Transitions and Singularities in Viscous Flow," Phys. Rev. Lett. {\bf 70},
3043 (1993); J. Eggers, ``Universal pinching of 3D axisymmetric
free-surface flow,'' Phys. Rev. Lett. {\bf 71}, 3458 (1993);
A.L. Bertozzi, M.P. Brenner, T.F. Dupont, and L.P. Kadanoff,
``Singularities and similarities in interface flows,'' in 
{\it Centennial Edition, Applied Mathematics Series}, L. Sirovich,
Ed. (Springer-Verlag Applied Mathematics Series, New York, 1993).}
\def\Lubpromise{}
\lref\rmembd{F. Brochard and J. Lennon, ``Frequency spectrum in the
flicker phenomenon in erythrocytes,'' J. Phys. France {\bf36} (1975)
1035
; M. Schneider, J. Jenkins, and W. Webb, ``Thermal fluctuations of
large cylindrical phospholipid bilayers,'' Biophys. J. {\bf45} (1984)
891; ``Thermal fluctuations of 
large quasi-spherical bimolecular phospholipid vesicles,''
J. Phys. {\bf45} (1984) 1457\pagin{--1472}
; S. Milner and S. Safran, ``Dynamical fluctuations of droplet
microemulsions and vesicles,'' Phys. Rev. {\bf A36} (1987) 4371\pagin{--4379}.
}
\lref\rBFM{R. Bar-Ziv, T. Frisch, and E. Moses, "Entropic expulsion in
vesicles",  preprint 1995 to appear in Phys. Rev. Lett.} 
\lref\rAskUdo{D. Marsh, {\sl CRC handbook of lipid bilayers,} (CRC,
1990) p.~166.}
\lref\HarHel{W. Harbich and W. Helfrich, {``Alignment and Opening of
Giant Lecithin Vesicles by Electric Fields,''} {Z. Naturforsch.}
{\bf34A} (1979) 1063\pagin{--1065}. Note however that these authors
considered a much lower frequency than the one of interest to us.}
\lref\rSeLi{U. Seifert and R. Lipowsky,  ``Morphology of vesicles'', in
{\sl Structure and Dynamics of Membranes}, ed. {R. Lipowsky and E.
Sackmann} (Elsevier, 1995).}
\lref\AbSt{M. Abramowitz and I. Stegun, eds., {\sl Handbook of
mathematical functions} (Dover, 1965).}
\lref\rOCMP{G. Oster, l. Cheng, H. Moore, and A. Perelson, ``Vesicle
formation in the Golgi apparatus,'' J. Theor. Biol., 141 (1989)
463\pagin{--504}. }
\lref\rfred{U. Seifert, ``Curvature-induced lateral phase separation
in two-component vesicles,'' Phys. Rev. Lett. {\bf 70} (1993)
1335\pagin{--1338}.}
\lref\rLeibinst{S. Leibler, ``Curvature instability in membranes,''
J. Phys. France 47 (1986) 507--516.} 
\lref\BMMS{R. Bar-Ziv, R. Menes, E. Moses, and S. Safran, ``Local
unbinding of pinched membranes,'' preprint 1995 to appear in
Phys. Rev. Lett.}
\lref\rMeSaEv{R. Merkel, E. Sackmann, and E. Evans, ``Molecular
friction and epitactic coupling between monolayers in 
supported bilayers,'' J. Phys (Paris) {\bf50} (1989) 1535.}%
\lref\CaHe{P. Canham , J. Theor. Biol. {\bf26} (1970) 61; W.
Helfrich, Naturforsch. {\bf28C} (1973) 693.}
\def\MSWDetc{\budtemp\MSWD}
\lref\budtemp{U. Seifert, K. Berndl, and R. Lipowsky, \cmp{``Shape
transformation of vesicles,''} Phys. Rev. {\bf A44} (1991) 1182;
L. Miao, B. Fourcade, M. Rao, M. Wortis, and R.K.P. Zia, ``Equilibrium
budding and
vesiculation in the curvature model of fluid membranes,'' Phys. Rev.
{\bf A43} (1991) 6843.}
\lref\MSWD{L.
Miao, U. Seifert, M. Wortis, and H.-G. D\"obereiner, \cmp{``Budding
transitions of fluid-bilayer vesicles,''} Phys. Rev. {\bf E49} (1994)
5389\pagin{--3407}.}%
\lref\SeLa{U. Seifert and S. Langer, \cmp{``Viscous modes of bilayer
membranes,''} Europh. Lett. {\bf23} (1993) 71\pagin{--76}; 
\cmp{``Hydrodynamics of membranes: the bilayer aspect and adhesion,''}
Biophys. Chem. {\bf49} (1994) 13\pagin{--22}.}
\lref\Plateau{J. Plateau, {\sl Statique experimentale
et theorique des liquides \cmp{soumis aux seules forces moleculaires}}
(Gautier-Villars, 1873).}
\lref\Safi{S. Chiruvolu, H. Warriner, E. Naranjo, S. Idziak; J. Radler,
R.                   Plano, J. Zasadzinski, and C. Safinya, \cmp{``A
phase of liposomes with
entangled tubular vesicles,''} Science {\bf266} (1994) 1222\pagin{--1225}.}
\lref\EvYe{E. Evans and A. Yeung, \cmp{``Hidden dynamics in rapid
changes of bilayer shape,''} Chem. Phys. Lipids, {\bf73} (1994)
39\pagin{--56}.}%
\lref\Tomo{S. Tomotika, ``On the instability of a cylindrical thread
of a viscous liquid surrounded by another viscous fluid,'' Proc. Roy.
Soc. Lon. {\bf A150} (1932) 322.}%
\lref\rNPS{P. Nelson, T. Powers, and U. Seifert, ``Dynamic theory of
pearling instability in cylindrical vesicles,''
Phys. Rev. Lett. {\bf74} (1995) 3384\pagin{-3387}.}
\lref\WvS{W. van\thinspace Sarloos, ``Front propagation into unstable
states,'' Phys. Rev. {\bf A37} (1988) 211\pagin{--229}.}
\lref\DeeL{G. Dee and J. Langer, ``Propagating pattern selection,''
Phys. Rev. Lett. {\bf50} (1983) 383\pagin{--386}; E. Ben-Jacob, H.
Brand, G. Dee, L. Kramer, and J. Langer, ``Pattern propagation in
nonlinear dissipative systems,'' Physica {\bf14D} (1985)
348\pagin{--364}; J. Langer, in {\sl Chance and matter,} ed.
J. Soulettie, J. Vannimenus,and R. Stora (North-Holland, 1987).}

\lref\rGallez{E. Gallez and W. Coakley, ``Interfacial instability at
cell membranes,'' Prog. Biophys. Molec Biol. 48 (1986) 155\pagin{--199}.}
\lref\rRothm{J. Rothman, ``Mechanisms of intracellular protein
transport,'' Nature {\bf372} (1994) 55\pagin{--63}.}
\lref\EvHe{E. Evans, Biophys. J.\rr{, ``Bending resistance and chemically 
induced moments in membrane bilayers,"} {\bf14} (1974) 923; W. Helfrich, \cmp{%
``Blocked lipid exchange in bilayers and its
possible influence on the shape of vesicles,''} Z. Naturforsch.
{\bf29C} (1974) 510.}
\lref\Evansc{E. Evans, Biophys. J. {\bf14} (1974) 923; Biophys. J.
{\bf30} (1980) 265.}%
\lref\SBZ{S. Svetina, M.B. Brumen, B. Zeks, ``Lipid bilayer elasticity
and the bilayer couple interpretation of red cell shape
transformations and lysis,'' Studia Biophys, {\bf110} (1985) 177.}%
\lref\Evansa{E. Evans, ``Entropy-driven tension in vesicle membranes
and unbinding of adherent vesicles,'' Langmuir {\bf7} (1991) 1900.}
\lref\Evansb{E. Evans, A, Yeung, R. Waugh, and J. Song, \cmp{``Dynamic
coupling and nonlocal curvature elasticity in bilayer membranes,''} in
{\sl Structure and conformation of amphiphilic membranes,} ed. R.
Lipowsky {\it et al} (Springer, 1992).}%
\lref\MiaSeif{U. Seifert, K. Berndl, and R. Lipowsky, \cmp{``Shape
transformation of vesicles,''} Phys. Rev. {\bf A44} (1991) 1182; 
L. Miao {\it et al.}, \cmp{``Equilibrium budding and
vesiculation in the curvature model of fluid membranes,''} Phys. Rev.
{\bf A43} (1991) 6843} 
\lref\Miaoa{L. Miao {\it et al.}, ``Equilibrium budding and
vesiculation in the curvature model of fluid membranes,'' Phys. Rev.
{\bf A43} (1991) 6843.}
\lref\Seifert{U. Seifert, K. Berndl, and R. Lipowsky, ``Shape transformation
of vesicles,'' Phys. Rev. {\bf A44} (1991) 1182.}
\lref\Bouasse{H. Bouasse, {\sl Cours de math\'ematiques
g\'en\'erales} (Paris, Delagrave, 1911).}
\lref\DeHeb{H. Deuling and W. Helfrich, \cmp{``The curvature elasticity of
fluid membranes: a catalog of vesicle shapes,''} J. Phys. (Paris) {\bf
37} (1976) 1335.}
\lref\ShSi{M. Sheetz and S. Singer, ``Biological membranes as bilayer
couples,'' Proc. Nat. Acad. Sci. USA {\bf 71} (1974) 4457.}
\lref\DeHe{H. Deuling and W. Helfrich, \cmp{``A theoretical explanation
for the myelin shapes of red blood cells,''} Blood Cells {\bf3} (1977)
713; \cmp{``The curvature elasticity of
fluid membranes: a catalog of vesicle shapes,''} J. Phys. (Paris) {\bf
37} (1976) 1335.}%
\lref\exphydra{D. Needham, private communication.}
\lref\Miaob{L. Miao {\it et al.}, ``Budding transitions of
fluid-bilayer vesicles,'' Phys. Rev. {\bf E49} (1994) 5389.}
\lref\Bozic{B. Bozic {\it et al.}, ``Role of lamellar membrane
structure in tether formation from bilayer vesicles,'' Biophys. J.
{\bf 61} (1992) 963.}
\lref\WHH{W. Wiese, W. Harbich, and W. Helfrich, ``Budding of lipid
bilayer vesicles and flat membranes,'' J. Phys. Cond. Mat. {\bf4}
(1992) 1647.}
\lref\Sveta{S. Svetina, A. Ottova-Leitmannov\'a, and R. Glaser,
``Membrane bending energy in relation to bilayer couples concept of
red blood cell shape transformations,'' J. Theor. Biol. {\bf94} (1982)
13.}
\def\Mich{}
\lref\Helfcz{W. Helfrich, ``Blocked lipid exchange in bilayers and its
possible influence on the shape of vesicles,'' Z. Naturforsch.
{\bf29C} (1974) 510.\optional{[invents ADE model; `the spontaneous
curvature of a bilayer around a vesicle is zero if the two halves
including the adjacent aqueous phases are in thermodynamic equilibrium']}}%
\lref\Tombook{P. Chaikin and T. Lubensky, {\sl Principles of condensed
matter physics} (Cambridge, 1994).}%
\lref\Sambook{S. Safran, {\sl Statistical thermodynamics of surfaces,
interfaces, and membranes} (Addison-Wesley, 1994).}%
\lref\Darcy{D. Thompson, {\sl On growth and form} (Cambridge, 1917).}%
\lref\Wimley{W. Wimley and T. Thompson, ``Transbilayer and
interbilayer phospholipid exchange in 
dimyristoylphosphatidylcholine/dimyristoylphosphatidylethanolamine large
unilamellar vesicles,'' Biochemistry {\bf 30} (1991)
1702.\optional{[estimate lipid flipflop times]}}%
\lref\MoBZ{R. Bar-Ziv and E. Moses, \cmp{``Instability and `pearling'
states produced in tubular membranes by competition of curvature and
tension,''} Phys. Rev. Lett., {\bf73} (1994) 1392; R. Bar-Ziv and E.
Moses, to appear.}%
\lref\MBpromise{R. Bar-Ziv and E.
Moses, to appear.}
\lref\OYH{Ou-Yang Zhong-can and W. Helfrich, \cmp{``Bending energy of
vesicle membranes,''} Phys. Rev. {\bf A39} (1989) 5280.}
\lref\HeSe{W. Helfrich and R. Servuss, ``Undulations, steric
interaction and cohesion of fluid membranes,'' Nuovo Cimento {\bf3D}
(1984) 137.}
\lref\EvRa{E. Evans and W. Rawicz, ``Entropy-driven tension and
bending elasticity in condensed-fluid membranes,'' Phys. Rev. Lett.
{\bf64} (1990) 2094.}
\lref\Bessis{M. Bessis, {\sl Living blood cells and their
ultrastructure,} (Springer, 1973) pp.~61, 170.}
\lref\Pouligny{B. Pouligny, private communication; see also M. Angelova, B.
Pouligny, G. Martinot-Lagarde, G. Grehan, and G. Gouesbet, ``Stressing
phospholipid membranes using mechanical effects of light,'' Prog.
Colloid Polym. Sci. {\bf97} (1994) 293\pagin{--297}, and in press.}
\lref\Thpriv{B. Thomas, private communication.}
\lref\Dober{H.-G. Dobereiner, private communication.}

\lref\SeSch{J. Selinger and J. Schnur, ``Theory of chiral lipid
tubules,'' Phys. Rev. Lett. {\bf71} (1993) 4091.}
\lref\Cates{
M. Cates, ``The Liouville field theory of random surfaces,'' Europhys.
Lett. {\bf8} (1988) 719.\optional{[6/88]}}%
\lref\DeTa{P.G. de\thinspace Gennes and C.  Taupin, J. Phys. Chem.
{\bf86} (1982) 2294.}

%
%
\lref\EvSk{E. Evans and R. Skalak, {\sl Mechanics and thermodynamics
of biomembranes}  (CRC Press, 1980).}%
\lref\WSSZ{R. Waugh, J. Song, S. Svetina, and B. Zeks, \cmp{``Local and
nonlocal curvature elasticity in bilayer membranes by tether
formation,''} Biophys. J. {\bf61} (1992) 974.}%
\lref\BSZW{B. Bozic, S. Svetina, B. Zeks, and R. Waugh, ``Role of
lamellar membrane structure in tether formation from bilayer
vesicles,'' Biophys. J. {\bf61} (1992) 963.}%
\lref\BrMe{F. Brochard-Wyart, J.-M. di$\,$Meglio, and D. Qu\'er\'e,
``Theory of the dynamics of spreading of liquids on fibers,'' J. Phys.
France {\bf51} (1990) 293.}
\lref\BrLe{F. Brochard and J. Lennon, ``Frequency spectrum in the
flicker phenomenon in erythrocytes,'' J. Phys. France {\bf36} (1975)
1035.}
\lref\Seidyn{U. Seifert, ``Dynamics of a bound membrane,'' Phys. Rev.
{\bf E49} (1994) 3124.}
\lref\Brugro{R. Bruinsma, ``Growth instabilities of vesicles,'' J.
Phys. II France {\bf1} (1991) 995.}
\lref\Strutta{Lord Rayleigh, ``On the instability of jets,'' Proc. Lond.
Math. Soc. {\bf10} (1879) 4.}%
\lref\Struttb{Lord Rayleigh, \cmp{``On the instability of a cylinder of viscous
liquid under capillary force,''} Phil. Mag. {\bf34} (1892) 145.}%
\lref\Goren{S. Goren, ``The instability of an annular thread of
fluid,'' ?? 309.}%
\lref\Liprev{R. Lipowsky, ``The conformation of membranes,'' Nature
{\bf349} (1991) 475.}
%
%
\lref\Messager{R. Messager, P. Bassereau, and G. Porte, ``Dynamics of
the undulation mode in swollen lamellar phases,'' J. Phys. France
{\bf51} (1990) 1329.}
\lref\bigchiral{P. Nelson and T. Powers, ``Renormalization of chiral
couplings in tilted bilayer membranes,'' J.  Phys. France II  {\bf3}
(1993) 1535.}

\lref\TCLWC{W. Cai and T.C. Lubensky, \cmp{``Covariant hydrodynamics of
fluid membranes,''} Phys. Rev. Lett. {\bf73} (1994) 1186.}



\tnewsec\Sintro{Introduction and Summary}

The study of artificial biomembranes has opened a window into the
machinery of real cells by giving us physical systems which are simple
enough to describe from first principles, yet complicated enough to
display lifelike behavior. The study of the equilibrium configurations
of closed bags of lipid bilayer\rr{s} (``vesicles'') is by now well advanced
\Liprev \rSeLi.
Most interesting biophysical phenomena, however, are not in
equilibrium, and  the study of the {\it dynamics} of vesicle shapes is
somewhat less developed. For example, budding and vesiculation \rRothm\ and
instabilities crucial for understanding adhesion \rGallez, are all
inherently dynamical processes.

To gain a physical understanding of non-equilibrium
membrane dynamics we must begin
with experiments in which some {\it physical} intervention brings
about a dynamical shape transformation.\foot{Experiments in which
a {\it biological} intervention is introduced (for example, mutating a
single protein) are comparatively abundant \rRothm.} Several such
techniques are by now well developed. Budding and other shape
transformations can be induced by adjusting temperature in closed
vesicles \MSWDetc.  Strings of beads can be induced by relieving
tension using micropipettes \EvRa, by sudden hydration of dry lipid
\exphydra, and by the incorporation of
a small object which is subsequently pulled away 
with laser tweezers 
\Pouligny. \kk{One can even create tension inside a vesicle by
polymerizing a stiff rod inside it~\rMiHo,
whereupon a cylindrical structure 
modulated with pearls can emerge~\ref\rfyg{D. Fygenson, private
communication.}.}

Recently Moses and Bar-Ziv have introduced a new class of experiments
in which laser tweezers act {\it directly} on single lipid bilayers \rr{\MoBZ}.
It is already  perhaps surprising that there should be any action at
all, since the optical absorption  of a single bilayer is so small.
As we will recall in Sect.~\Fphys\ below, we argued in \rNPS\ that the effect
of the
tweezers is to induce a {\it tension} $\Sigma$ in the membrane
proportional to the laser power. Our argument relied only on basic
electrodynamics, and while it does not give the precise value of the constant of
proportionality, it does imply that the effect of the laser is
extremely simple; unlike other more mechanical probes, the possibility
of complicated parasitic effects seems minimal. More importantly,
laser tweezers provide for the first time the fast, delicate, and
highly localized probe needed to understand membrane dynamics in
detail. For example, while beaded tubes have been seen in some of the
other techniques mentioned above (see Sect.~2), a unique feature of
the tweezer experiment is the controlled ability to excite a
``small-amplitude'' (quasilinear) peristaltic modulation.

We recall the basic results of \MoBZ\ in Sect.~2 below. The most
intriguing phenomenon for us was the fact that when laser tweezers are
applied to previously stable long cylindrical vesicles, they excite an
instability ending with a ``string of pearls'' state.  The initial
wavelength of the instability is  $\lambda=2\pi R_0/k$, where
$R_0$ is the initial cylinder radius and the
dimensionless wavenumber is typically $k=0.8$.
Remarkably, a
purely {\it local} excitation due to the laser trap creates (in a certain
regime) a {\it uniformly} modulated state, which invades the initial
cylindrical region at a roughly {\it constant velocity} of about
$v_f\sim$30\micron/sec. This is not the sort of behavior we usually
associate with pulling a stretchy object in a viscous fluid, and so we
have a very interesting dynamical system.

The pearling phenomenon bears a superficial resemblance to the
instability of a cylindrical column of water in air, studied in the
classical works of Plateau and Rayleigh \Plateau\Strutta\Struttb. In
this situation
surface tension destabilizes the cylinder, since the same volume per
unit length can be contained with less surface area as a string of
spherical droplets. As in other pattern-forming systems, a competition
ensues: the lowest surface-to-volume ratio comes from a small number
of large droplets, but the formation of such a state is kinetically
suppressed because it requires the motion of water (a conserved
quantity) over long distances.
For a macroscopic system the water may be treated
as inviscid (high Reynolds number); Rayleigh calculated that the
fastest-growing instability set in at wavenumber
$k=0.70$\ \Strutta. 
At micron scales,
however, we are in a regime of {\it low} Reynolds number; here
Rayleigh found the fastest-growing instability to be at $k=0$\
\Struttb, totally different from what is seen in \MoBZ.

Of course the situation studied in ref.~\MoBZ\ is not a thread of 
liquid surrounded by air. Long ago
Tomotika considered a {\sl two-fluid model,} in which a column of one
viscous fluid is initially immersed in another with a certain positive
surface tension \rr{\Tomo}. One could imagine that these two fluids correspond to
the interior and exterior water, so that both viscosities are equal.
Specializing Tomotika's result to this
case, one finds the fastest-growing
mode to be at $k=0.56$. Unfortunately this simple picture is
not obviously adequate for the pearling problem, nor indeed is its
predicted value for $k\max$ ever observed; experimentally the initial
wavenumber is always a bit larger, and then increases still further
with time. What is missing?

For one thing, the two-fluid model neglects the presence of a material
{\it object} between the inner and outer fluids, namely the membrane,
except insofar as the latter somehow communicates tension induced at
the trap to the entire surface. In particular Tomotika imposed
continuity of the shear stress $T_{\rho z}$ across the boundary, while
a material object located there will in general be capable of exerting
tangential forces and hence causing a discontinuity in $T_{\rho z}$.

A more realistic treatment of membrane dynamics requires
a {\sl three-fluid model,} in which a
{two-dimensional} fluid (the bilayer) separates two
three-dimensional fluids (the water). The ``tension'' on the membrane
can then more precisely be regarded as minus the 2d {\it pressure}
$\Pi$ of the intermediate fluid, which in turn emerges from the
appropriate hydrodynamic equations. The key early works on membrane
dynamics used models of this type to study equilibrium fluctuations~\rmembd.
In the pearling experiment the laser takes the system {\it far} from
equilibrium by suddenly imposing
a {\it boundary condition} on $\Pi$. In other words,
the laser trap is regarded as a lipid {\it reservoir} located at $z=0$ in
cylindrical coordinates, whose chemical potential suddenly jumps from
zero to a negative value as the laser is switched on. 
\optional{Later (\frictsect) we will
elaborate the model by resolving}

The idea that the laser effectively induces a tension $\Sigma$, and
that a Rayleigh-type instability ensues, was first proposed by Moses
and Bar-Ziv \MoBZ. We substantiated this picture of the laser action
in \rNPS, showed how to estimate $\Sigma$ from the laser power, and
\optional{worked through a partial treatment of the four-fluid model to find}
found the fastest-growing mode $k\max=0.68$, at the low end of the observed values.\foot{Actually we
quoted $k\max=0.65$ due to an error which
we will correct in eqn.~\fecons\ below.}
\optional{We found that the simpler
three-fluid model actually gave very similar answers, and we explained
why.}

The most serious lacun\ae\ in the analysis of ref.~\rNPS\ were that
(a) we neglected the {\it propagating} character of the pearling
instability, assuming instead that the tension was everywhere a
constant, and (b) we neglected the fact that the lipid fluid is {\it
conserved}. These points are related, of course. For a shape
transformation to propagate outward from the trap, leaving behind a
stationary shape, lipid must constantly be transported away from the
moving front and into the trap. Assumption (a) gave us no possibility at
all of predicting the front {\it velocity}, a readily accessible
experimental \rr{quantity}. 

In this paper we will present a more detailed analysis addressing
these and many other points. After reviewing the experimental facts in
Sect.~2, we will build up a physical picture of the pearling system in
Sect.~\Fphys. In particular we will show that assumption (b) above is a good
approximation during the time regime of the experiment, so that
pearling is essentially the invasion of a saturating front into a
uniform linearly-unstable state. To get to the point as quickly as
possible we will construct in Sect.~\Fvsm\ a very simple form of the
three-fluid model which captures most of the essential physics without
lengthy formulas. We then analyze the front propagation in
Sect~\Ffront\ using the marginal stability criterion (or ``MSC'') \DeeL\WvS\ and
show how it gives qualitatively correct results for $v_f$ and $k$ when
applied to our simple model. As we will recall, the MSC allows one to compute the front
velocity using only the {\it linearized} dynamics.

To get quantitatively accurate predictions we will construct a more
realistic model in Sect.~\Fexact. The reader may wish to skip this
rather technical section. In it we first solve the hydrodynamics
exactly, then
resolve  the bilayer into its two monolayers,
to get a {\sl four-fluid model}. The motivation for this is that
dynamical friction between the two leaves of the bilayer has been
shown to be quite significant for a related class of membrane problems
\Evansb\SeLa\EvYe, though as we will see the actual effect on our
answers will be slight.  Again using the
MSC, we  first find that the initial  wavenumber is $k_0=0.80$, which
agrees with experiment somewhat better than the result of \rNPS. In
particular the initial wavenumber is insensitive to
changes in the laser power or tubule radius, as observed. Secondly
we get a front velocity $v_f$  equal to
$0.06\Sigma/\eta$ where $\eta$ is the viscosity of water, another
prediction in rough agreement with experiment. While this
prediction isn't
precise due to our imprecise knowledge of $\Sigma$, it does explain
why the magnitude of $v_f$ is observed to be so much smaller than the
\mm{natural velocity scale $\Sigma/\eta$.} Moreover,
the linear dependence of the initial front
velocity on the laser
power should be verifiable in the future.

Since the MSC uses only the linearized equations it must assume,
rather than proving, the existence of a uniformly-propagating
front. One could take the phenomenological attitude that such behavior
is observed experimentally, but this and other qualitative facts
should emerge from the 
solution of the full nonlinear
equations. In Sect.~\Fnon\ we introduce a third model, with most of
the simplifications of the first, simple model (Sect.~\Fvsm) but
retaining the nonlinear structure of the elasticity.
We show the results of
numerical studies that indicate that it indeed  supports propagating front solutions, 
and that it produces stationary pearled structures like those seen in experiment. 
More details will appear elsewhere\Lubpromise.

Appendix A contains a glossary of symbols used for physical constants,
and their values.

\optional{In fact we will argue that (b)  and that (a)
too is a reasonable guide for getting the wavelength selection, though
neither of these is obvious (to us) {\it a priori}. In fact our
present analysis refines our prediction of Also
We will  argue that this
velocity may be computed by invoking the marginal stability criterion
. We will have to work a bit to justify the applicability
of this formalism to our case\foot{Actually a full justification
requires a detailed analysis of the nonlinear structure of the
equations \WvS. In this paper we will give careful analysis only to
the linearized theory, making only qualitative remarks about the
nonlinear structure. Our attitude is that the existence of a
propagating front is an experimental fact, though we will sketch why
it is to be expected from the physical model we develop. In any
case   the quantitative results thus obtained are fairly successful.%
} (see \MSCsect), but it is worth the
effort: 
In this way }

Granek and Olami were the first to attempt a realistic treatment of pearling
without the simplifying assumptions (a), (b) above \GrOl. While our
physical picture and conclusions are  different from theirs, we
are indebted to them for emphasizing these issues. We are also
grateful to Moses and Bar-Ziv, who first suggested to us that marginal
stability might be applicable to this problem. Finally, others have suggested
that the laser could have other effects besides inducing tension, for
example effectively inducing a bilayer asymmetry (spontaneous
curvature) \GrOl\ref\Svet{S. Svetina, private communication.}. We
don't see how this could happen, and in any case we will see that  no
such effect is needed
in order to explain the observed phenomena. 

\tnewsec\Sexp{Relevant Experimental Facts}\nobreak
Briefly the observed phenomena of interest to us are as follows \MoBZ.
Initial preparation of the system yields stable,
nearly straight cylinders up to hundreds of microns long, anchored
at both ends by large globules of lipid. Each tube consists of a
single bilayer of DMPC or DGDG, with water on the inside and outside.
The tubules are {\it polydisperse}, with initial radii $R_0$ between
0.3--5\micron. The high temperature used
($45^\circ$C) implies that the membrane is in its pure fluid state.
Initially the system is somewhat
flaccid, as seen from visible thermal undulations and the fact
that the tubes are not quite straight.

Application of a laser spot localized to $\sim0.3$\micron\ produces a
dramatic  transformation to a stationary ``peristaltic'' 
figure, \ie\ a cylindrical
shape with radius $r(z)$ at first varying roughly sinusoidally with distance
$z$ from the trap. There is no evidence for any initial time delay
between laser illumination and the onset of the instability.
Greater laser power is required for larger-radius
tubules, but nothing seems to depend  on the {\it length} of the tube,
so long as the trap is initially many radii from the ends.
The shape transformation {\it propagates} outward
from the laser trap, with a well-defined velocity $v_f$ which varies
between experimental trials but is typically about 30\micron/sec and
roughly constant for at least a couple dozen wavelengths. Remarkably,
after a very short illumination the shape transformation continues to
propagate {\it after} the laser is shut off, leading to a uniform,
small-amplitude peristaltic shape. Longer excitation leads to the
eponymous pearled state.
Sometimes tubes intersect each other; in this case the shape
transformation can actually cross from one tube to the other upon
reaching the intersection~\MBpromise. 

Once formed, the peristaltic shape has
a well-defined initial wavelength \rr{$\lambda$} which is uniform over many
microns. \rr{The dimensionless initial wavenumber $k_0=2\pi R_0/\lambda$
is always found in the range 0.64--1.0, and is typically about $0.8$
(ref. \MoBZ, fig.~1),
whatever the initial radius $R_0$}. After prolonged tweezing
some buildup of lipid becomes visible at the point of application of
the laser. As the modulation grows
more pronounced, $k$ grows from $k_0$ to become slightly
greater than 1 and deviations from a simple sinusoidal profile become
pronounced.  The modulated state is {\it tense}: visible thermal
fluctuations are suppressed and the tube draws itself straighter than
initially.

We pause to contrast the above phenomena with other related results.
Deuling and Helfrich studied aged red blood cells, which had different
interior and exterior fluids; in this situation it is reasonable to
invoke a spontaneous curvature $c_0$ expressing the chemical asymmetry
of the bilayer environment. In our case there is no such asymmetry; as
we will recall in Sect.~\FFineq\ below, the fact that each leaf of the
bilayer is initially in equilibrium with a common lipid reservoir then
implies that $c_0=0$ \EvHe.

Evans and Rawicz, and recently Pouligny,
have rapidly formed thin tethers coming from large vesicles, respectively by
adjusting  the internal pressure or by pulling out a small inclusion
in the membrane \EvRa\Pouligny. Such tethers sometimes contain pearls.
Also lower-temperature experiments, where in-plane order is important,
can give pearled shapes \Thpriv.
All of these
experiments have in common that the formation of pearls is coterminous
with the formation of  the tethers themselves, and small-amplitude
sinusoidal modulations are not seen. Finally, experiments in {\it
mixed} surfactant systems have shown {\it equilibrium} peristaltic shapes
\Safi. This is presumably due to shape-composition coupling, which is
known to lead to an instability towards an asymmetric bilayer \rLeibinst\rfred\
and thence to the mechanism of \DeHe.

\tnewsec\Sphys{Physical Picture} 
In
Rayleigh's problem, the tension was a {\it material parameter}
characterizing the air-water interface, and hence trivially constant.
In our case we argued in the Introduction that tension is instead a
dynamical variable, and so its spread must be self-consistently
determined along with the change of shape.
Our main objective in this section is to justify the assumption
\MoBZ\rNPS\ that
tension initially spreads so rapidly that it effectively becomes
constant and uniform as soon as
the laser is turned on\foot{As mentioned earlier,
Granek and Olami first studied this issue \GrOl.}. Thus the spread of tension is {\it
not} what limits the speed of initial propagation of the shape pulse. 
Our arguments in this section will all be rather crude. We can only
emphasize that the field of membrane dynamics is not yet fully
developed, and buttress our arguments whenever possible with the
observed phenomenology.

In this section we will for simplicity assume axial
symmetry everywhere. We will return to this point in \elasticsect. Thus all
scalars are functions of radius $\rho$, distance $z$ from the trap,
and time $t$; all velocity vectors have vanishing azimuthal component.

\tnewsubsec\SSineq{Initial equilibrium}
We begin with a discussion of the initial equilibrium, before turning
on the laser.

Initially some preparation protocol has created long cylindrical
vesicles constrained to stretch between the terminal blobs and in
thermal equilibrium with them. We will speculate as little as possible
about the nature of the blobs. Since they are far from the
illumination spot, their only role is to determine the nature of the
initial equilibrium, not the subsequent fast dynamics.
We will simply assume that the blobs furnish lipid reservoirs and that
both leaves of the bilayer membrane are in equilibrium with these
common reservoirs, with a chemical potential for exchange close to
zero.

With this assumption we find that initially the membrane has tension
close to zero, consistent with the initially observed thermal motion
\MoBZ. We also get that the  spontaneous curvature $c_0=0$, since in a
chemically symmetric bilayer $c_0$ can only arise as the {\it
difference} in chemical potentials between the two constituent
monolayers\foot{See Appendix \FAade. The situation is quite different in {\it closed}
systems, such as vesicles of spherical topology. Here indeed the
initial preparation leads to a new parameter describing vesicles, the
fixed {\it area difference} between the two monolayer leaves \MSWD.}
\EvHe. $c_0=0$ is consistent with the observed absence of any
preferred initial tubule
radius \MoBZ.
Nonzero spontaneous curvature can lead to {\it equilibrium}
unduloid shapes \DeHe\OYH, but as we mentioned earlier these are not
observed in the experiments in question. Moreover, these ``Delaunay
surfaces'' begin with an initial wavenumber $k=1$\rr{,} larger than what is
observed (see Appendix B).

\mm{The physical boundary condition imposed by the terminal blobs
will not matter in our analysis, but for concreteness suppose that
they seal off the tubule, imposing fixed volume. This constraint}
then creates a physical {\it pressure}
difference, with a very small negative constant pressure inside the
tube $p_0=-\kappa/2R_0^{\ 3}$
(and $p=0$ outside), where $\kappa$ is the usual bending modulus. $p_0$
balances the tendency of a cylinder to increase in
diameter to reduce its bending energy. One can either solve the volume
constraint explicitly and substitute into the bending energy (as in
\MoBZ\rNPS), or else regard $p_0$ as  a Lagrange multiplier (as in
\OYH) to see that then the cylindrical shape is stable to all small
perturbations. This is the initial equilibrium state.

\tnewsubsec\SSlaser{Laser action} 
Next we recall our model of the laser action from ref.~\rNPS.
When the laser comes close to the membrane, nothing
happens: local heating is not important. When the laser spot {\it
touches} the membrane, it pulls material in by the dielectric
effect. While it is hard to calculate the exact tension so
induced, we may easily estimate it as follows\foot{We thank R.
Bruinsma for suggesting this estimate.}: the applied laser power of
$\sim50$mW, spread over a spot of diameter $0.3$\micron, corresponds
to an energy density in vacuum $\CE$ of $2.3\cdot10^4$\erg cm$^{-3}$.
Taking
the dielectric contrast between water and lipid at this frequency to
be of order $\delta\epsilon=0.23$
\ref\Isrbook{J. Israelachvili, {\sl Intermolecular and surface forces}
(Academic Press, 1992).}%
, we see that when a lipid
molecule falls into the trap displacing water we gain an energy
\ref\JDJ{J. D. Jackson, {\sl Classical electrodynamics} (Wiley, 1975)
pp. 107, 160.}{}
$\sim\CE\dl
\ep\cdot a_0D$, where $a_0$ is the area of the molecule's head and $D$ is
its total length. Taking $2D\sim40$\AA, we get that each unit of
bilayer area sucked into the trap yields an energy gain of
$\Sigma\sim2\cdot10^{-3}$\erg\cmmt.

Actually  this figure is surely an
overestimate. If the action of the laser is really to pack more lipid
into the trap, there must be an offsetting cost per unit area to fold
it up or otherwise put it into a more compact
configuration than a single nearly-flat bilayer. \kk{(For more details see
\rBFM.)} Still,
we see that the trap generates a tension well in excess
of the critical value~\MoBZ\ $\Sigma\crit\sim{\kappa\over
R_0^2}\sim1.2\cdot10^{-4}$\erg \cmmt\ needed to trigger shape transformations
(see \elasticsect\ below),
where $\kappa\sim0.6\eem{12}$\erg\ is the bending stiffness of DMPC
bilayers and we took $R_0=0.7$\micron\ for illustration.
Let $
\sigma\equiv\Sigma{R_0}^2/\kappa$
denote the dimensionless
tension; thus
\eqn\esigmaest{\sigma\equiv\Sigma {R_0}^2/\kappa\sim2\ee1\quad,}
and larger for larger tubules. \kk{Such large tensions are
qualitatively corroborated by other experiments involving laser-induced
expulsion of vesicles \rBFM.}
\let\edfsig=\esigmaest

One could imagine that the details of how the bilayer gets packed into
the optical trap could effectively create {\it different} boundary
conditions for the chemical potential of each layer and hence induce a
sort of symmetry breaking, a spontaneous curvature for the layer. For
example, \mm{tiny vesicles} could get pinched off
preferentially from the
inner layer. We will resist such speculations, as we have no
theoretical or experimental evidence in favor of such a mechanism, nor
will such a spontaneous curvature be needed to understand the phenomena.
We will also assume that the laser creates no major disruption to the
integrity of the bilayer, \ie\ the membrane remains impermeable on the
time scale (tenths of a second) in question\foot{This amounts to
assuming that the volume is effectively clamped by dissolved
macromolecules. It is reasonable since large laser-induced tension can pressurize
vesicles with no significant loss of interior volume, even long
after the laser is shut off \rBFM. Indeed far
greater tensions than those considered here can be applied
mechanically with no observed loss of 
interior volume nor other breakdown in the bilayer model \EvRa.%
}, and transfer of lipid from one
leaf
to the other (``flipflop'') also remains too slow to be of interest,
as in unstressed membranes \Wimley. 

\kk{Further support for our model of the origin of pearling by
laser-induced tension comes from the observation that {\it mechanical}
tension applied suddenly to a cylindrical vesicle (by dragging an end
attached to a movable pipette) creates a pearling instability similar
to the one seen with the laser \MBpromise. Whichever excitation one
uses, when it stops} the tension reverts slowly to zero, thermal
fluctuations resume, and the tubule relaxes back to its initially
stable cylindrical shape~\MoBZ.

\tnewsubsec\SSinprop{Initial tension propagation} 
As we mentioned in the Introduction, the induced tension $\Sigma$ is
properly to be regarded as a {\it boundary condition} on the {\it
pressure} $\Pi$ of the lipid fluid\foot{The pressure of a 2d fluid has
dimensions of tension.}:
$$\Pi(z=0,t>0)=-\Sigma\quad.$$
The sudden introduction of a gradient of $\Pi$ causes the lipid to
move toward the laser spot; this motion in turn stretches the
membrane, causing the tension to spread outward. We will first argue
that this initial spreading of tension is very fast, and 
essentially complete long before shape changes have had a chance to
begin. Hence in {\sl this subsection only}  let us at first neglect shape changes
altogether and ask what happens \mm{in a linearized analysis} when we
begin to pull on a cylinder of 2d fluid, then later check the
self-consistency of this picture. Since the shape is fixed and the
water is incompressible, there can be no net flow of water across a
cross-section of the tube.

As the bilayer begins to move toward the
trap,  it entrains the surrounding water. Focusing our
attention  on the interior volume, this entrainment sets up a velocity
profile $v_z(\rho,z,t)$. For this estimate we will use Poiseuille (or
``lubrication'')
approximation, where
all gradients in $z$ are assumed much smaller than those in $\rho$.
(We will also temporarily neglect the {\it exterior} fluid.) The incompressible
conservation law $\nabla\cdot v=0$ then implies that the radial
velocity is small, $v_\rho\ll\vz$, and by the equations of motion the
pressure $p$ is constant across the cross-section of the tube.
Solving the remaining equation of motion $\eta\nabla^2
v_z=\nabla_z p$ in this
approximation then gives us the usual parabolic velocity profile of Poiseuille flow,
\eqn\ePoisa{v_z(\rho,z,t)=f(z,t)\rho^2 -g(z,t)\quad,}
where $f$ is related to the pressure gradient and $g=\half R_0^{\
2}f$ because there can be no net flow of water down the tube.

The membrane itself is a compressible 2d fluid. Its velocity $\vt$ must
match that of the water: $$\vt=v_z(R_0)= \half R_0^{\ 2}f$$
For its equation of motion, we can neglect the 2d viscosity of the lipid
on these scales \SeLa\ and write the force-balance equation as
$$\nz\Pi(z,t)=T_{z\rho}(R_0,z,t)=-\eta\pa_\rho\vz|_{R_0}=-2\eta R_0f
(z,t)\quad,$$
\kk{where $T_{ij}$ is the 3d stress tensor of water.}
Gradients in the velocity must in turn affect the {\it density} of
lipid, by another conservation law. We write the lipid density as $\phi=
\phi_0(1+\chi)$ where $\phi_0$ is the equilibrium density and $\chi$
is the areal strain; then
\eqn\elipconsa{{\dd\over\dd t}\left[\dd z\,2\pi R_0\phi_0(1+\chi)\right]=-\dd
z\cdot\nz
\left[2\pi R_0\vt\cdot\phi_0(1+\chi)\right]\quad.}
To close the equations we need the constitutive relation
$$\Pi=K\chi\quad,$$
where $K$ is the 2d bulk modulus of the lipid layer.

Combining we find
\eqn\epdispersion{\pd{\Pi}{t}={KR_0\over4\eta}\nabla_z^2\Pi\quad,}
whose solution $\Pi(z,t)$ indeed spreads rapidly until it is
essentially equal to the boundary condition $-\Sigma$ throughout the
observed tens of microns. To estimate how rapidly, we need the value of $K$.
We argue in Appendix~\FAmodulus\ that $K$ is effectively much smaller than its
``bare'' value\foot{Were we to use the bare value $K_0$, we would have
to include inertial effects in the above derivation, leading to a
different dispersion relation from \epdispersion\ \SeLa, but with the same
qualitative conclusion: tension spreads more quickly than shape change.}
 $K_0\sim1.4\ee2$\erg\cmmt; instead we will argue for
$K\eff\sim10^{-1}$\erg\cmmt. Even so, $K\eff\gg\Sigma$. Combined with
\esigmaest\ we get the hierarchy of scales
\eqn\ehierarchy{\kappa R_0^{\ -2} \ll \Sigma \ll K\eff\quad,}
which we will use repeatedly. 

The large modulus means that
the tension profile will
rapidly outrun any front of shape transformation traveling at $v_f\le
\Sigma/\eta$ (in fact we will find in Sect.~\Fexact\ that $v_f$ is much smaller than
this). For technical reasons the observed
region must contain the laser spot, so propagation is not observed for
values of $L$ larger than about 50\micron. 
Scaling \epdispersion\ we see that at a distance
$L$ from the trap the tension approaches its saturation value in a
time of order $t_L\sim L^2\eta/K\eff R_0$
. Taking $L=50$\micron\ gives $t_L\sim4\eem2\,$sec, about one video
frame, and much faster than the time $L/v_f\sim 2\,$sec needed for the
observed front propagation.

We conclude that after a very short time  the tweezers create
a uniform, tense, stretched state of the lipid in the region of
interest. If the tension $\Sigma$ is great enough, this state will be
everywhere linearly unstable to shape perturbations, just as in the
Rayleigh instability. Of course it remains to show that a uniform
stretched state will indeed change shape \via\  a uniformly propagating
front as assumed above.

\tnewsubsec\SSfronta{Front propagation} 
We  have just argued  that initially the tension spreads rapidly, before
the shape has had a chance to change. \rr{To see} what happens to
tension as the instability progresses, \rr{w}e will continue to assume that
this proceeds by the propagation of a front, leaving behind a
stationary modulated state. \rr{Later, this is shown to be a}
self-consistent picture.

The energy which drives our instability comes from delivering lipid to
the trap. But where does this
lipid come from? Certainly very little comes from {\it stretching} the
bilayer, since we just asserted that its bulk modulus is much greater
than the applied tension. Instead it comes from a shape
transformation, in which the initial cylindrical shape gets replaced
by something with less area (and hence fewer lipid molecules) per
length. It may at first be hard to see how such a transformation could
propagate at a constant velocity. After all, lipid must constantly be
transported to the trap, entraining water as in the previous
subsection and incurring viscous dissipation. Much as in Poiseuille
flow through a pipe, shouldn't we expect a decrease of pressure $|\Pi|$ as the
front moves away from the trap?

To address the question, consider first a case in which the
cylinder gets converted at a moving front to a modulated 
shape, with a change of its area per unit length from $2\pi R_0$ to
$2\pi R_0(1-\alpha)$. The front is located at $z=L$, so to advance it
at a velocity $v_f=\dd L/\dd t$ lipid must flow toward the trap at
velocity $\vt=\alpha \vf$.

The laser does work on the membrane at a rate $2\pi
R_0\alpha\vf\cdot\Sigma$. This energy is lost due to viscous
dissipation both at the front, and everywhere from $z=0$ to $L$. Since the
front is only a few times $R_0$ in length, we can estimate the former
loss as $\eta{\vf}^2\pi R_0$. The latter loss will be mainly due to
entrainment of water.\optional{\foot{We note in passing that this entrainment
will also set up a pressure gradient in the water, which could perhaps
drive the later migration of pearls back toward the trap \MoBZ. This
migration continues after turning off the laser, so some sort of
energy storage mechanism is needed to explain it.}} By the remarks in
the previous subsection, the entrainment of water creates a shear
$\pa_\rho\vz\sim \alpha\vf/R_0$
in a volume $\pi {R_0}^2L$, for a loss rate of $\eta
\alpha^2{\vf}^2\pi L$. Comparing the total loss rate to the power
input, we find that
\eqn\evfest{\vf\propto{\Sigma\over\eta}{2\alpha\over\zeta+\alpha^2L/R_0}\quad.}
Here $\zeta$ collects various order-one factors neglected in our rough
estimates. We see that
indeed the front cannot propagate at a constant velocity forever. But
$\vf$ {\it can} stay roughly constant until $L>\zeta\alpha^{-2}R_0$,
and the regime in which this happens is characterized by the fact that
drag on the membrane behind the front is negligible. \kk{(The late-time
asymptotics have been studied by Granek and Olami \GrOl.)}

What is the fractional area loss $\alpha$? Suppose first that the
modulation saturates behind 
the front at a small amplitude, so that its radius is (\tfig\fcyl)
\eqn\edfr{r(z,t)=R_0(1+u(z,t))}
with $|u|\ll1$. Requiring that the new shape have the same {\it
volume} per unit length as the old (no global transport of fluid), we
find that $\alpha$ must be $\CO(u^2)$, so that $R_0/\alpha^{2}$ is
enormous. We will argue that this case is relevant for the case of a
short laser pulse, when full pearls never form at all.

\ifigure\fcyl{Modulated cylinder.}{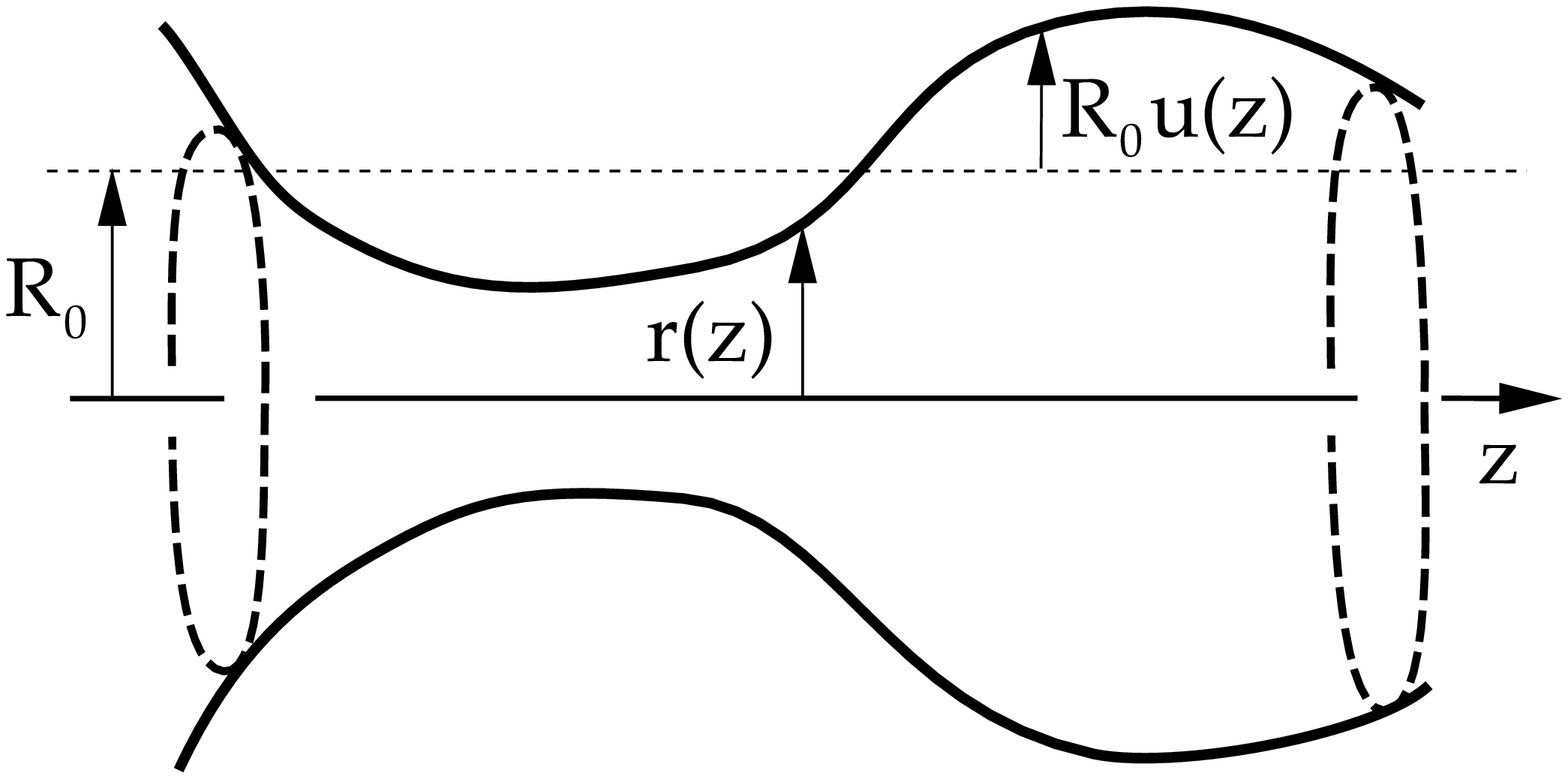}{1.5}

If on the other hand the laser is kept on for a long time, we expect
the final state to be of large amplitude $u\sim1$, so one might worry
that $\alpha\sim1$. But we know a lot about the final state: it must
be a string of pearls on thin tethers. After all,
neglecting bending stiffness the system minimizes area at fixed
volume, leading to spherical pearls. Kinetic limitations prevent the
pearls from all merging; instead we expect a periodic structure with some
wavelength $\lambda$. The effect of the curvature stiffness is to
prevent the \mm{tethers from shrinking to zero size.}
Since a thin cylinder
of radius $R_1<R_0$ and length $\ell$ has curvature energy $\lfr\kappa2
(2\pi R_0\ell)\bigl(\lfr1{R_1}\bigr)^2$,
$R_1$ will stabilize at a
nonzero value. Indeed if the tension is large (as we have argued),
$\Sigma/\Sigma\crit\gg1$  where $\Sigma\crit=\kappa/R_0^2
$, then we
expect tethers of {\it small} radius, as seen.\foot{\kk{In fact the
minimization gives a tether radius of $\sqrt{\kappa/\Sigma}
$, times 
some constants of order unity, which is qualitatively correct. There
is thus no need to invoke spontaneous curvature to explain tether
formation.}}  In this case we
can neglect the volume and area of the tethers altogether.

Thus volume conservation says that a string of pearls of radius $R_2$,
separated by thin tethers of length $\ell$ (so that the wavelength
$\lambda=2R_2+\ell$), must have volume per length
${4\pi(R_2)^3\over3\lambda}=\pi
(R_0)^2$. The area per unit length is then just $4\pi(R_2)^2/\lambda$. 
For the wavelengths $\lambda\sim2\pi R_0$ characteristic of the observed
nonlinear regime, this corresponds to a fractional area loss $\alpha$
of just 10\%. In other words, {\sl the constraint of volume
conservation has kept $\alpha$ small}. If we carry over the estimate
leading to \evfest\ (even though now the amplitude is not small), we
see that once again $\alpha^{-2}$ is very large and since $\zeta$ is
$\CO(1)$, we
can neglect the drag along the length of the tube for the initial
propagation. Any gradient of the tension must be due to the drag, and
moreover we argued in Sect.~\SSinprop\ that the tension starts out
uniform, so the continued smallness of the drag means that the tension
stays uniform in the initial propagation regime.

The conclusion of this subsection and the previous one is that \kk{in
the regime of interest} we may
reasonably model the effect of the laser as generating a {\it constant
uniform tension} on the membrane, just as was assumed in
\MoBZ\rNPS. Lipid cannot literally disappear locally, but in the
observed initial regime we may act as though this were the case. In
other words, the pearling system behaves as if it were suddenly {quenched}
into a {\it uniform unstable state}. This is the sort of situation in which
we expect front propagation, as we recall in Sect.~\Ffront\ below.

As mentioned earlier, our conviction doesn't rest solely on these crude arguments,
but also on the experimental fact that indeed a front does form and
propagate at constant velocity, at least initially~\MoBZ. Later in
Sect.~\Fnon\ we will see the front formation again in the numerical solution
to a simplified model.

\tnewsubsec\SSother{Other assumptions} 
Here we collect other assumptions we are making.

Our axisymmetric assumption \edfr\ could be generalized to allow
$u(z,\varphi,t)=\sum_m u_m(z,t)\ex{\rmi m\phi}$.  The peristaltic mode
is $m=0$. In \elasticsect\ we will see that the $m\ge 2$ modes are stable.
There remains the $m=1$ mode, which corresponds to {\it wandering} of
the cylinder centerline. The $m=1$ mode is always soft, or critical,
since it corresponds to the broken symmetry of displacement normal to
the tube. As we mentioned, such wandering is indeed observed
initially.  However we will neglect it 
since (at least in the linearized analysis) it decouples from the
interesting peristaltic mode.

We also assume in our hydrodynamic analysis no-slip boundary
conditions between the fluid bilayer and the surrounding water. This
is a standard assumption and no evidence exists to make us relax it.
However, the two leaves of the bilayer {\it can} slip relative to each
other, with a friction coefficient measured in the
experiments of Evans and collaborators \Evansb\EvYe. We will
incorporate these effects in Sect.~\Fexact\ below.

\subsec{Summary}
Let us summarize this long section. We have given a qualitative, intuitive
argument for a physical picture in which the laser creates a nearly
uniform, tense state, both before and during the subsequent shape
changes. We {\it assumed} that the latter proceeded by the
propagation of a front, which then must initially move at a velocity
$\vf$ considerably smaller than the dimensional combination
$\Sigma/\eta$ (see eqn.~\evfest), as observed. We still need to show that
this is the case and get a precise formula for  $\vf$.

Since front propagation in general leaves behind a state of wavenumber
$k_0$ different from the fastest-growing mode $k\max$ \DeeL\WvS, we
will also need to revisit our estimate of ref.~\rNPS\ to see what
happens to that prediction. \optional{We will also replace the
approximate hydrodynamics used there by something more accurate
(\threesect) and
ultimately incorporate the effects of bilayer stiffness and
friction (\frictsect).} We do this for a simple model in Sect.~\Ffront, then
for a more accurate model in Sect.~\Fexact.

\tnewsec\Svsm{Very Simple Model}

To get as quickly as possible to the heart of the matter we first
develop a simple truncated model without a lot of algebra.
In {\it this tutorial model only} we will make some
expedient additional assumptions, each of which we will justify or
improve in the sequel. The assumptions are:
\itemitem{{\it i)}} As in \rNPS, we will temporarily
neglect the propagating character of the modulation until
Sect.~\Ffront. Instead we will 
just find the wavenumber of the fastest-growing Fourier mode.
Our goal here is
to show in the simplest possible way how tension leads to wavenumber
selection.
\itemitem{{\it ii)}}
We will neglect
altogether the bending stiffness of the membrane as well as all
bilayer effects until Sect.~\Fexact.2.  This will prove to be a good approximation if the
applied tension $\Sigma$ greatly exceeds the threshold value for
instability, as we have argued is the case (see eqn.~\ehierarchy).
\itemitem{{\it iii)}} We will neglect the
finite compressibility of the membrane. This is legitimate as long as
$\Sigma$ is much  smaller than the effective elastic modulus $K\eff$
(see eqn.~\ehierarchy\ and Appendix~\FAmodulus). \optional{Thus we
suppose the hierarchy \ehierarchy.
Introducing the dimensionless tension $\sigma\equiv\Sigma{R_0}^2/\kappa$,
we thus have $\sigma\gg1$.}
\itemitem{{\it iv)}} We will use lubrication approximation as in
\ePoisa. This is certainly {\it not} justified, since the wavenumber $k\max$
of the mode we will find is not much smaller than unity and hence $z$
derivatives are not much smaller than $\rho$ derivatives, but it will
make our equations very simple. Similarly we will neglect the exterior
fluid altogether.

We argued that as soon as the laser switches on, very quickly a nearly
uniform tension $\Sigma$ appears. Volume conservation then implies
that the pressure inside jump to a constant positive value,
$p_1=\Sigma/R_0$ (the Laplace pressure of a cylinder under tension) to
prevent the tube from collapsing. As the tube starts to change shape,
described by the small quantity $u$ in \edfr,
\mm{the Laplace pressure $\Delta p\lap(z,t)=-\Sigma\cdot2H$ gets new contributions from the
change in the
mean curvature $H$ \OYH:
\eqn\epressa{\Delta p\lap={\Sigma\over R_0}\bigl[1-u-{R_0}^2\nz^2u\bigr]+\CO(u^2)\quad.}

Since we temporarily neglect the exterior fluid,
the interior fluid must move in such a way as to give the pressure
$p(r(z,t),z,t)=\Delta p\lap$, which we just computed.} To linear order
we again  have the fluid flow
\ePoisa, where now $f,g$ are functions of $z$ and time. We can no
longer fix $g$ by demanding that the net flow of water across any
cross-section vanish, since now the tube is getting fatter in some
places and thinner in others. Similarly we need to modify the lipid
conservation law. Generalizing \elipconsa\ to variable
radius but fixing the density gives
\eqn\elipconsb{{\dd\over\dd t}\left[\dd z\,2\pi r(z,t)\right]=-\dd z\cdot\nz
\left[2\pi r(z,t)\vt\right]
}
where the boundary velocity $\vtz=\vz(r(z,t),z,t)=fr^2-g$. Thus linearizing
in small quantities we get $\dot u=-\nz [f{R_0}^2-g]$. Similarly,
conservation of the interior water gives
\eqn\econswater{{\dd\over\dd t}\left[\dd z\,\pi r(z,t)^2\right]=-\dd z\cdot\nz
\left[\int(2\pi\rho\dd\rho)\vz\right]\quad,
}
or $\dot u=-\nz[\lfr14 f{R_0}^2-\half g]$. Comparing \elipconsb\ we
find\foot{Note that in \rNPS\ we mistakenly set the boundary velocity
to zero instead of fixing it with \elipconsb--\econswater. The formula
for $k\max$ below is however unaffected.}
\eqn\eudota{\dot u=\half {R_0}^2\nz f\quad.}
Since we are linearizing, let us take $u(z,t)$ to be a single Fourier
mode:
\eqn\eusin{u_k(t)\cos(kz/R_0)\quad.}
The fluid equation of motion $\eta\nabla^2 v_z=\nabla_z p$, together
with \epressa\ then gives
(neglecting $z$ derivatives compared to $\rho$ derivatives) $4 f=
-{\Sigma\over\eta R_0} \nz(1-k^2) u$, or (see \tfig\fdisp)
\eqn\esimplest{\dot u={\Sigma\over8\eta R_0} k^2(1-k^2) u\quad,
}
so that the mode with the largest growth rate $\dot u/u$ is at
$k\max=1/\sqrt 2$. Note that the tension needed to get an appreciable
growth rate increases with tubule radius, as observed.

\ifigure\fdisp{Exact and lubrication approximation dispersion
relations, neglecting stiffness and bilayer structure.  \rr{Circles
indicate fastest-growing modes.}}{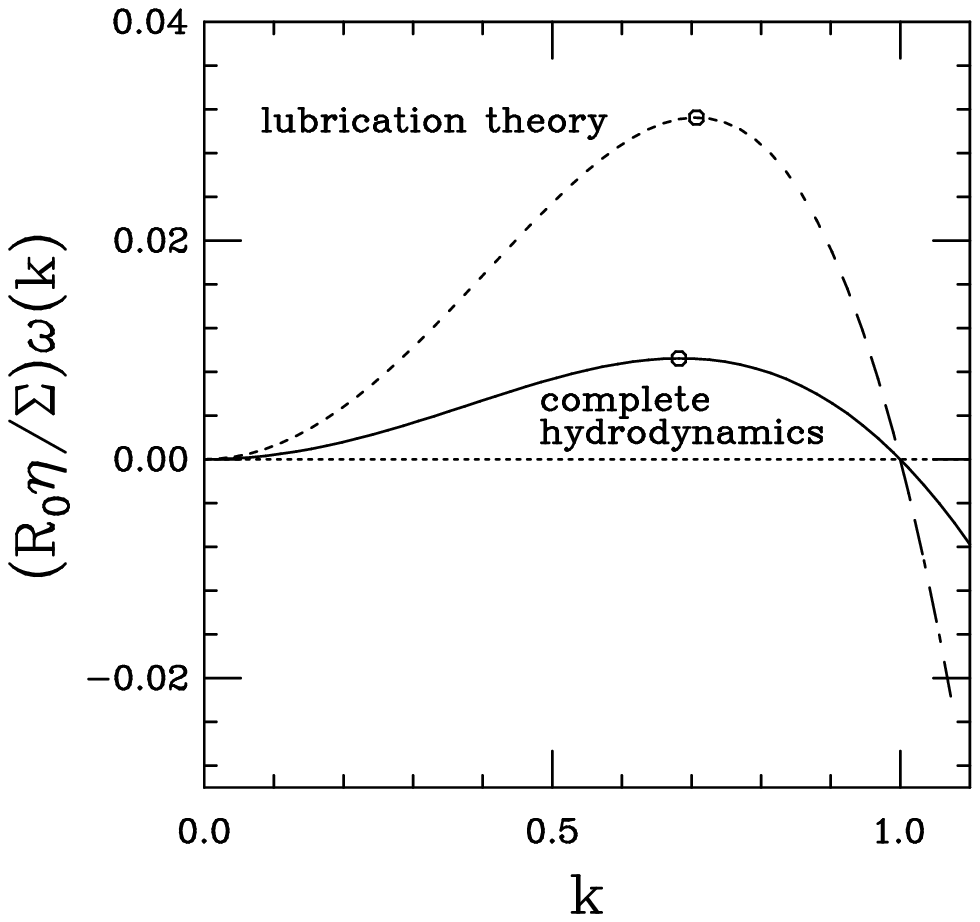}{4.5}

Note the structure of the dispersion relation \esimplest\ ; the growth
rate is the product of a term ($\propto(1-k^2)$) arising from
the perturbation to the pressure in \epressa, while
the additional overall factor of $k^2$ represents two spatial
derivatives coming from the hydrodynamics.  
The first is due to the conservation of fluid volume, and
the second from the usual Poiseuille-flow relation between fluid flux
and pressure gradients.  In the more refined hydrodynamic theory 
derived below the general structure is retained, in the sense that the
growth rate is the product of a hydrodynamic factor and a pressure term, 
but the former has a more complicated wavevector dependence than $k^2$ as 
a consequence of the
details of flow in cylindrical geometry.

\tnewsec\Sfront{Velocity and Wavelength Selection} 

We saw in our very simple model that tension destabilizes a
cylinder. More generally when we incorporate bending stiffness in
Sect.~\Fexact\ we'll 
see the same thing, with a finite threshold $\Sigma\crit$.
We now want to know what
becomes of such an unstable state.

We argued in Sect.~3 that in the regime in question our problem could
be regarded as a {\it uniform, quenched} unstable state: the membrane
elastically transmits the laser-induced tension everywhere. The
uniform unstable state changes its conformation until it saturates in
one of two ways. For brief tweezing, we suppose that the laser pulls
out about 1\% of the area (see Appendix~\FAmodulus), then shuts off;
even when the laser stops pulling the shape transformation proceeds to
an amplitude such that the
projected area is reduced by about 1\%, \ie\ $u\sim\sqrt{0.01}$. This
fits with the observation of a small-amplitude peristaltic shape which
continues to propagate after shutoff. For sustained tweezing, there is
no such limitation, and the shape transformation continues until
checked by the curvature energy of the thin tethers, as we argued in
Sect.~\SSfronta.

In either case a uniform quenched unstable state rolls off a potential
hill to saturate nonlinearly. In this
class of problems one typically finds that the instability
proceeds via  the propagation of a {\it front} from an initial
disturbance into the unstable medium.\foot{In the present case the
laser itself provides the initial disturbance; for another example see the
photographs in \BMMS.} The question of the precise
conditions under which an equation has such a front solution are
ticklish and require a more reliable knowledge of the full nonlinear
equation than we possess (for example, at nonlinear orders the effects
of thermal fluctuations are surely more subtle than the picture in
Appendix~\FAmodulus). However, if we take the existence of the front as an
empirical fact, we can use the analysis of \WvS\ to determine its
main properties solely from the linearized dynamical equations.

The shape is governed by some nonlinear, translation-invariant
equation in $u(z,t)$ 
We
suppose a family of solutions whose envelopes are functions of $z-vt$,
for various values of the velocity $v$. Only one of these solutions
can correspond to the actual observed front; thus we seek the
dynamically selected value $v^*$. The selected value must at least be
stable to perturbations, in an appropriate sense which we now
recall.

Suppose the front moves from negative to positive $z$ at velocity
$v^*$ and consider the
leading edge, where $u$ is still small. We need to examine the effect of small
perturbations $\dl u$. In this region we may use the linearized
equations, so that superposition holds and the perturbation belongs to
a family $\dl u\sim \exp(\omega(q) t-qz)$ for various values of a
complex wavenumber $q$, where $\omega(q)$ is the dispersion relation of
the linearized equation. The envelope of this blip is then $\exp(t\Re\omega
-z\Re q)$, which is itself translating at uniform velocity
$v_q=\Re\omega/\Re q$. Of this family of perturbations, not all are
dangerous to the stability of the front. A wavenumber $q$ is dangerous
only if
\itemitem{{\it i)}} $\Re q>0$, so that this is a leading edge solution,
\itemitem{{\it ii)}} $\Re\omega>0$, so that it is growing, {\it and}
\itemitem{{\it iii)}} $v_q\ge v^*$.

\noindent Without {\it (iii)}, the blip just gets left behind by the front and
never gets a chance to destabilize it. If any complex $q$ meets all
these conditions we will call the front unstable in its comoving
frame.

Of course the assumed front solution itself corresponds at its leading
edge to one particular linear solution with some $q^*$, and so
\eqn\ewvsa{v^*=\Re\omega_*/\Re q^*}
where $\omega_*=\omega(q^*)$. Let us now consider values of $q$ very
close to $q^*$, $q=q^*+a+\rmi b$ for small $a,b$. Then we find
$$v_q=v^*\left[1+
{a\over \Re\omega_*}\bigl(\Re\omega'_* - v^*\bigr)
- {b\over\Re \omega_*}\Im\omega'_*
\right]\quad,$$
where $\omega'_*=\left.{\dd\omega\over\dd q}\right|_{q^*}$. Since
$a,b$ were arbitrary, and $q^*$ obeys {\it (i), (ii)} above, we can
always satisfy {\it (iii)} as well unless
\eqn\ewvsb{\Im 
\left.{\dd\omega\over\dd q}\right|_{q^*}=0
\ ,\qquad v^*
=\Re\left.{\dd\omega\over\dd q}\right|_{q^*}\quad.}
The three equations \ewvsa--\ewvsb\ are necessary conditions for the
linear stability of a propagating front in its comoving frame \WvS. We will refer to them as
the ``marginal stability criterion for $v^*$'' or simply ``the MSC'',
and use them to fix the three real variables
$v^*,\  \Re q^*,\ \Im q^*$. There is a large literature on
experimental tests of pattern selection via the MSC. Important early
studies include refs.~\rXMSC. 

We can also use \ewvsa--\ewvsb\ to determine the selected wavenumber $q_0$
of the stationary pattern behind the front.  In the comoving frame of the
front, the saturated pattern is continuously created at the front and
moves rigidly with velocity $-v^*$.  Nodes are created in the leading
edge of the front at a rate $\Omega={\rm Im}(\omega_*-q^*v^*)$.
We can interpret $\Omega$ as the flux of nodes moving toward the
saturated pattern; if nodes are not created or destroyed as they pass
into the non-linear region, then we must have $q_0v^*=\Omega,$ or
\eqn\ekzero{q_0={\rm Im}(\omega_*-q^*v^*)/v^*}
for the wavenumber of the pattern \DeeL.  Note that $q_0$ is generally
different from both the fastest-growing mode $q\max$ and from $q^*$.


Let us apply \ekzero\ to
our very simple model.
We found in   Sect.~\Svsm\ that $\omega(k)={\Sigma\over{8\eta
R_0}}k^2(1-k^2)$, eqn.~\esimplest. Extending this formula to complex
values of $k$ and applying \ewvsa--\ekzero,
we find $k_0\equiv q_0 R_0=.77$, \rr{a decay length of $3.8R_0$, and}
$v^*=.20{\Sigma\over\eta}\sim 4\cdot10^2{\rm \mu 
m/sec}$. In fact we will see in Sect.~\Fexact\ that our more accurate calculation of $k_0$ is
quite close to the very simple model, while our prediction for $v^*$
will be much smaller.
The selected wavenumber $k_0$ is purely
geometrical since we can form no length scale from tension and viscosity;
in particular $k_0$ is independent of laser power. 

Of course the very
simple model is still rather crude, even with the refinement we just
made. We will upgrade it to a more 
accurate model in Sect.~\Fexact. Since the algebra is a bit involved,
the reader may want to skip this section and pass directly to the
simulation results in Sect.~\Fnon.
\optional{ Here we see that $k_0$
is in better agreement with the experimental values than $k\max=.71$.
Our prediction for $v^*$ is less precise since $\Sigma$ is not well-known,
although we can say that $v^*$ increases with laser power and is independent
of tubule radius.}

\tnewsec\Sexact{Complete Linear Model}

Now that we have some confidence from our very simple model, we need
to address some of the oversimplifications listed in
Sect.~\Svsm. First we will replace the lubrication approximation with
the exact  solution to the Navier-Stokes equation. Then we resolve the
bilayer structure of the membrane to get the four-fluid model
mentioned in the Introduction.

\tnewsubsec\SSbessel{Hydrodynamics: Exact linear theory} 

Now we go beyond the lubrication theory and
use the full linearized hydrodynamic equations for the interior and exterior
fluids. 
We begin \kk{with Stokes's observation} that the axial symmetry we
have assumed makes 
our problem effectively two-dimensional.  This, along with the
incompressibility of the water, 
\eqn\einc{\pd{v_\rho}{\rho}+{ v_\rho\over \rho}
+\pd{v_z}{z}=0\quad,} 
allows us to write the
interior and exterior water velocity in terms of 
the cylindrical stream function $\psi$:
\eqn\estr{v_\rho={1\over\rho}{\pd\psi z}\ , \qquad v_z
=-{1\over\rho}{\pd\psi \rho}\quad.}
The Navier-Stokes equations are now two equations in two unknowns, $\psi$
and the pressure $p$.  We eliminate the latter by differentiating 
$\eta\nabla^2\vec v_\rho=\pa_\rho p$ with respect to $z$, differentiating $\eta\nabla^2
v_z=\pa_z p$
with respect to $\rho$, and subtracting the two resulting equations to
obtain 
\eqn\eNSstr{\Bigl(\nabla^2-{2\over\rho}{\pa\over{\pa\rho}}\Bigr)^2\psi=0\quad.}

In Appendix \FAlambda\ we find the stream function $\psi$ required by a single mode of shape
distortion $u(t)\exp{i k z/R_0}$, enforcing the no-slip boundary condition and
lipid conservation.  This yields $v$ and $p$ in terms of
$\dot u$, and ultimately the stress tensor $T^+_{\rho\rho}$ inside
and outside of the tube which determines the normal force balance equation 
to relate $u$ and $\dot u$, {\it i.e.} to determine the growth
rate $\dot u/u=\omega(k)$.  As mentioned in the Introduction,  our
boundary conditions differ from 
\Tomo\ and \GrOl, where the shear stress $T_{z\rho}$ is taken to be continuous 
across the surface of the cylinder.
This boundary condition is appropriate for two viscous fluids that meet at
an interface \Tomo; in our case, however, there is a  material object at the
interface that supplies whatever forces are necessary to ensure lipid
conservation and the no-slip condition.%
\foot{In Sect.~\FFcyl\ we also address the
effect of the two-dimensional membrane viscosity and show that it is
dominated by the traction of the water. \optional{
Moreover, the large value of the membrane's bulk modulus implies that
the traction will only slightly compress the bilayer.
The lipid motion due to this compression will be small, entraining little water
and thus making a negligible contribution to the hydrodynamic dissipation.}}

The results for the growth rate have the form that generalizes \esimplest\ 
by the introduction of a more complicated ``dynamical factor" $\Lambda(k)$,
\eqn\eGR{\omega(k)={\Sigma\over R_0\eta}\Lambda(k)(1-k^2)\quad}
with $\Lambda(k)$ given by eqn.~\fekinfac\ of Appendix~\FAlambda.
\tfig\fkinfac\ compares this function  with the lubrication
approximation $\Lambda_{\rm lub}=k^2/8$.
Our dynamical factor $\Lambda (k)$ differs from the one 
given in \rNPS\ due to the correction mentioned below eqn.~\elipconsb;
however the $\Lambda (k)$ of \fkinfac\ leads to a $k\max$ of $.68$
(see \fdisp), very close to the value in  \rNPS.

\ifigure\fkinfac{Dynamical factor as a function of $k$, \rr{with fastest-growing
modes indicated.}}{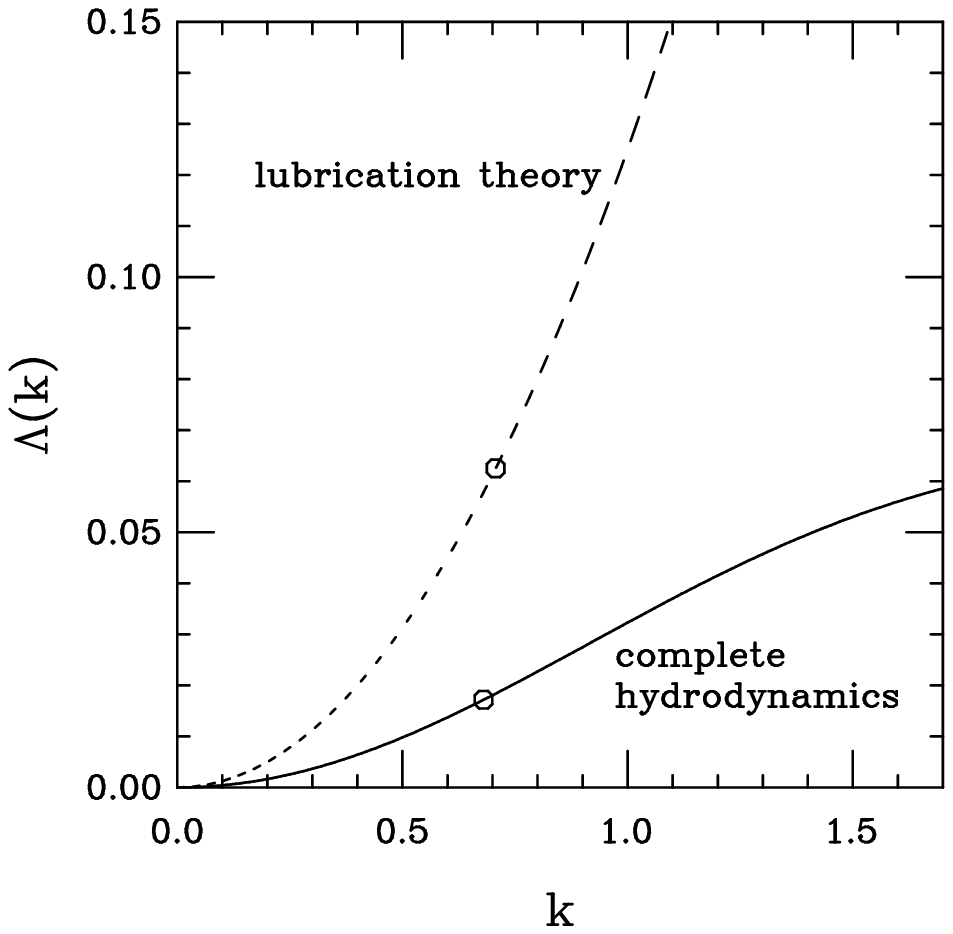}{4.5}

We see that for all wavenumbers the exact dynamical factor lies below
that given by lubrication theory.  This implies slower growth of
instabilities given the same material parameters $\Sigma$, $\kappa$,
and $\eta$ (see \fdisp).  This increased drag relative to the lubrication approximation
is sensible; the lubrication model neglects some components of the
velocity gradients and hence
underestimates the rate of viscous dissipation.

\optional{Within the three-fluid model we use the
growth rate \eGR\ with the dynamical coefficient $\Lambda(k)$ in the
MSC results \ewvsa--\ewvsb\ .  A numerical solution yields 
\eqn\elubMS{k_0=0.75 \ \ \ \ {\rm  and} \ \ \  v^*=0.06 {\Sigma\over \eta}\quad .}
{\df ??The value of $k_0$ is almost indistinguishable from the lubrication theory
result, and thus in equally good agreement with the experimental value $k\max=.71$.}
Note that the front velocity is
indeed much slower than $\Sigma/\eta$, consistent with our discussion
of \evfest. \optional{ Here we see that 
Our prediction for $v^*$ is less precise since $\Sigma$ is not well-known,
although we can say that $v^*$ increases with laser power and is independent
of tubule radius.}

Our dynamical factor $\Lambda (k)$ differs from the one given in \rNPS\
due to the correction mentioned in Sect.~??;
however the $\Lambda (k)$ of
\fkinfac\ leads to a $k\max$ of $.68$, indistinguishable from that of
\rNPS.}
Inserting the growth rate \eGR\ into the MSC \ewvsa--\ewvsb\ and solving
the equations numerically yields
$k_0=.75$ and $v^*=.06\Sigma/\eta$. We note that the front velocity is
indeed much slower than $\Sigma/\eta$, consistent with our discussion
of \evfest.

We see that taking the
propagating character of the instability into account has
slightly improved the agreement with the experimentally observed $k_0$. Using
our estimated value for $\sigma$, eqn.~\esigmaest, we also get the front
velocity  $v_f\sim100\ {\rm \mu m/sec}$. \mm{Since our value for
$\Sigma$ is an overestimate we get $v_f$ squarely in the
observed range.}

\optional{These results show that by taking into account the
propagating character of the instability we have
significantly improved the agreement with the observed $k_0$. Using
our estimated value for $\sigma$, eqn.~\esigmaest, we also get the front
velocity  $v_f\sim100\ {\rm \mu m/sec}$. \mm{Since our value for
$\Sigma$ is an overestimate we get $v_f$ squarely in the
observed range.}}


\tnewsubsec\SSbilayer{Bilayer model} 

\optional{Within the assumption of a rapidly equilibrated tension, we may now view
the membrane shape evolution as a gradient flow tending to minimize
the sum of energetic contributions from elasticity and tension.  
For tubular vesicles that energy is a functional of the position
vector ${\bf r}$ and in the absence of any
spontaneous curvature takes the form
\eqn\esimpleE{F[{\bf r}]=\int\dd S\left[\Sigma + {\kappa\over2}H^2\right]
\quad.}
Here $\kappa$ is the bilayer stiffness, $\dd S$ is the area element of the
bilayer, and $H$ is the mean curvature.  
From this elastic energy we derive a pressure difference across 
the membrane as the normal component of the functional derivative of $F$
with respect to the vector ${\bf r}$ that traces out the surface,
\eqn\efuncder{\Delta p= 
-{1\over \sqrt{g}}\hat {\bf n}\cdot{\delta F\over \delta {\bf r}}= 
\Sigma H -\kappa\left(\nabla^2 H +{1\over 2}H^3-2HK\right)\quad,}
where $g$ is the determinant of the metric tensor and $K$ is the
Gaussian curvature.
We may think of this as a generalization to the case of elastic surfaces of the 
bare Laplace pressure difference due to tension alone.
For future use note that
cylindrically symmetric shapes described by a radius $r(z)$ have mean  
and Gaussian curvatures given by
$$\eqalign{H&={r_{zz}\over {(1+r_z^2)^{3/2}}}-{1\over {r(1+r_z^2)^{1/2}}}\cr
K&= -{r_{zz}\over {r(1+r_z^2)^2}}~.}$$
Likewise, for these shapes the metric and covariant laplacian are
$$\eqalign{g&= 2\pi r\sqrt{1+r_z^2} \cr
\nabla^2&= {1\over r \sqrt{1+r_z^2}}{\partial\over \partial z} 
{r\over \sqrt{1+r_z^2}}{\partial\over \partial z}~.}$$
For uniform tubular states of radius $R_0$ we see from this that the balance 
of forces yields an equilibrium radius $R_0=(\kappa/2\Sigma)^{1/2}$, as discussed
in the previous section.

In the remainder of this and the following section we shall focus our attention
on the linearization of this pressure difference around a uniform cylindrically
symmetric shape, but in Section 8 return to address its full nonlinear form in
the context of a hydrodynamic model for pearling.

} 

We now need a more realistic model of the bilayer. For one thing, we
have so far neglected bending stiffness, which as we saw is eventually
crucial to stabilize the tethers between pearls. But stiffness effects
are comparable to effects involving the differential stretching of the
two leaves of the bilayer \EvHe\MSWD, so we must include these as well.
This means introducing more continuum degrees of freedom,
the densities $\phi^\pm$ of lipid molecules per unit  area in each
monolayer. 

\optional{To get started let us look more closely at the elasticity theory of
membranes. Elasticity theory is a phenomenological description of
extended bodies in equilibrium, in
which the many degrees of freedom corresponding to individual
molecules are replaced by a few continuum variables corresponding to
broken symmetries. In the case of fluid membranes, the broken symmetry
is translation normal to the surface, the corresponding variable is
{\it shape}, and the relevant terms of the elastic energy were studied
by Canham and Helfrich \CaHe. To introduce {\it dynamics}, in general
one must include a few more continuum degrees of freedom,
corresponding to conserved quantities. In our case these quantities
are the number of lipid molecules $N^\pm$ in each layer, and the corresponding
variables are the densities $\phi^\pm$ of molecules per unit  area.}

In fact one must sometimes retain $\phi^\pm$ even in equilibrium problems, for
example in closed systems where the numbers of molecules are
separately constrained. 
This situation leads to the ``area-difference elasticity'' model
studied for example in ref.~\MSWD\Mich. We can simply quote their
intermediate formula for the elastic energy:
\eqn\eADE{F[H,\phi^\pm]=\int\dd S\left[{\kappa\over2}(2H)^2+{
K_0\over4}\sum_\pm
\left({\phi^\pm\over \phi_0}-1\right)^2
+\Sigma\right]
\quad.}
(For completeness we rederive this formula in Appendix \FAade.) Here
$\kappa$ and $K_0$ are the {\it bilayer} stiffness and compression
moduli, and  $\dd S$ is the  area element of the
midplane. $H$ is the mean curvature of the bilayer
midplane; in our convention $H=-1/2R_0$ for a cylinder.
We measure the lipid density $\phi^+$ at the neutral surface of the
outer monolayer, assumed a distance $d$ away from the bilayer
midplane, and similarly $\phi^-$. See \tfig\fbilayer. Thus the
parameters of the model are $\kappa,\ K_0,\ d$,  and the full bilayer
thickness $2D$
(see Appendix A).  Since $D$ and $d$ are much smaller than any other length scale in
the problem, we will work to leading nontrivial order in them, as indeed
we have already done to get \eADE.

\ifigure\fbilayer{Schematic of the bilayer showing various variables.}{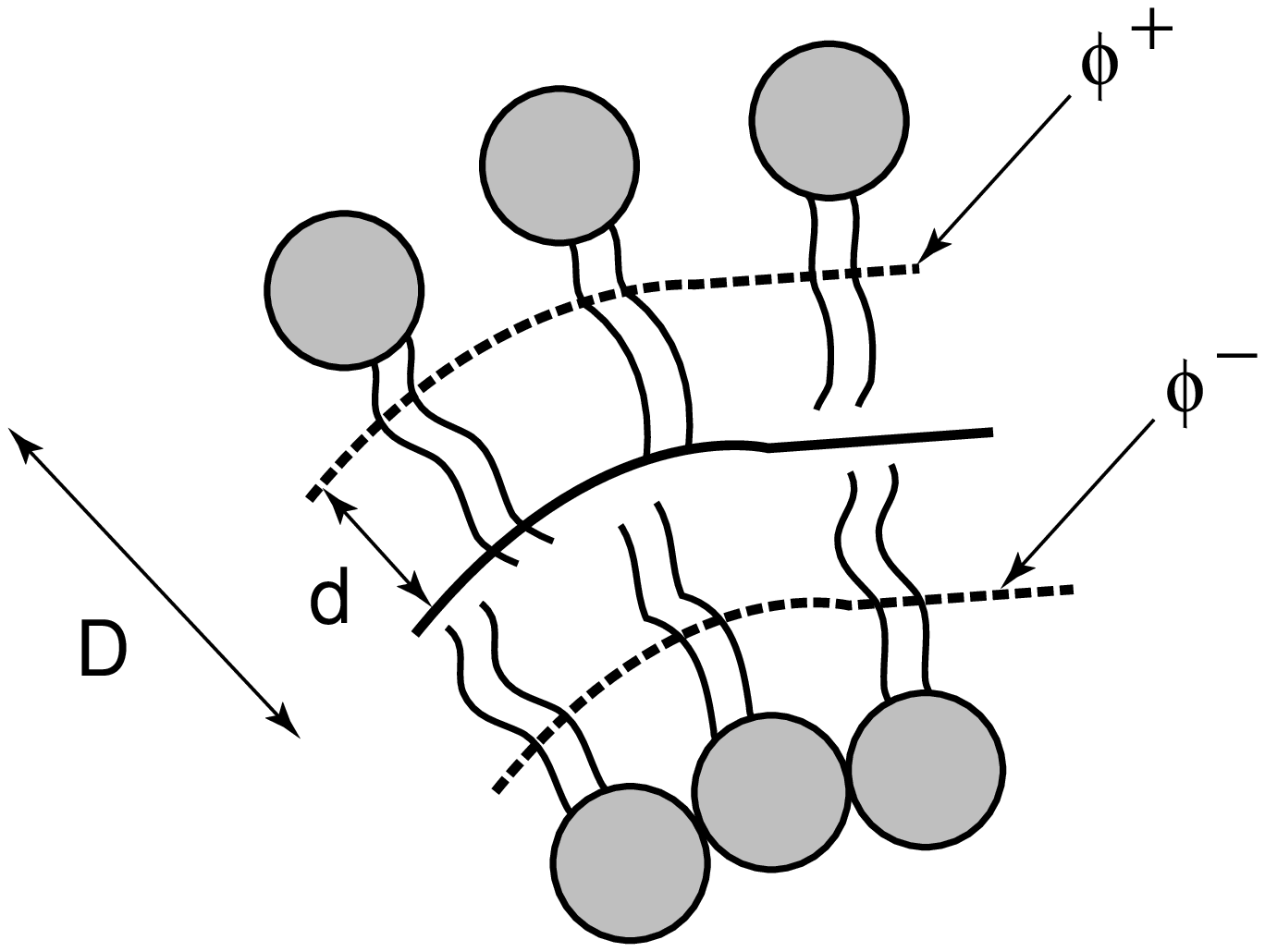}{2.5}

We begin in equilibrium at zero tension. Here the densities
$\phi^\pm$ take their preferred value $\phi_0$. Actually we will find it useful
to recast our equations in terms of the densities referred not to the
monolayer neutral surfaces, but to the bilayer midplane. If we imagine
each molecule casting its shadow on the midplane, the new density
variables $\chi^\pm$ are the  \mm{excess} density fluctuations of those
shadows\foot{In \rNPS\ we called these variables $\rho^\pm$.},
relative to the equilibrium densities $\phi_0^\pm$ projected to the
midplane of the unperturbed surface. Due to the
curvature, $\phi_0^\pm$ are no
longer equal, but instead  $\phi_0^\pm=\phi_0(1\mp 2H_0 d)$,
\rr{where $H_0$ is the equilibrium curvature}.
Thus  we have $1+\chi^\pm=\phi^\pm(1\mp 2Hd)/\phi_0^\pm$, or
\eqn\edfpsi{\phi^\pm/\phi_0=1+\chi^\pm\pm 2(\delta H)d
\quad,}
where
$\dl H=H-H_0$ is the change of curvature. Substituting in \eADE\ above
yields \rNPS
\eqn\eFone{F[H,\chi^+,\chi^-]=\int\dd S\left\{
{\lfr\kappa2}(2H)^2+
\lfr {K_0}4\bigl[(\chi^++2(\dl H)d)^2+(\chi^--2(\dl H)d)^2\bigr]
+\Sigma\right\}\ ,
}
Were we to consider
fluctuations about  {\it flat} membranes, we would set $H_0=0$; then
\eFone\ reduces to the formula in \SeLa.

Later we will want to rephrase \eFone \ in terms of the mean density
and density difference, and rescale:
\eqn\edfps{\bar\chi=(R_0/d)(\chi^++\chi^-)\
,\qquad\hat\chi=(R_0/d)(\chi^+-\chi^-)\quad.}
We argued in Sect.~3 that the mean density adjusts quickly to its
preferred value $\bar\chi=0$, since only hydrodynamic drag obstructs
this, and initially it is not effective. We cannot be quite so sure
about  density {\it difference}, however. The two monolayers are like a
pair of polymer brushes in contact, and so the  friction coefficient
between the two layers can be (and is) enormous \WSSZ\EvYe. 
Leaving $\hat\chi$
free, we then finally obtain the elastic energy
\eqn\eFtwo{\eqalign{F[H,\hat\chi]=\int\dd S\biggl\{&
\lfr\kappa2(2H)^2
+\lfr{K_0d^2}2 (2H-2H_0)^2\cr
&+\lfr{K_0}2\Bigl(\bigl(\lfr d{2R_0}\bigr)^2\hat\chi^2+\lfr{2d^2}{R_0}
      (H-H_0)\hat\chi\Bigr)
+\Sigma\biggr\}\ .\cr
}}

To gain a physical feeling for \eFtwo, consider
two limiting cases. If the interlayer friction is
small (or time scales long), we can minimize \eFtwo\ with respect to
$\hat\chi$ also, and recover once again the Canham-Helfrich model.
If the friction is very large (or time scales very short)
then $\hat\chi$ cannot change at all
from its initial value of zero, and we drop the third term of \eFtwo.
For flat membranes, $H_0=0$ and the first two terms combine to give
the effective increase of stiffness found in \SeLa.

\subsec{Cylinder perturbations}

Let us instead specialize \eFtwo \ to the case of a cylindrical initial
state and a small sinusoidal perturbation of $u$ and $\chi$ of wavenumber
$k$. Working to quadratic order
in fluctuations of shape and density difference one has \OYH
$$\eqalign{H=&\half {R_0}\inv\left[
-1 + u + {R_0}^2\bnabla^2 u - u^2\right] \cr
&+\half{R_0}\left[ - \half (\bnabla u)^2 - 2u\bnabla^2 u + (\nz u)^2
+ 2u\nz^2u  \right] \quad ,\cr
\dd S=& ( R_0\,\dd z\dd\varphi)\bigl(1+u +\half{R_0}^2(\bnabla u)^2\bigr)\quad.\cr}$$
In these formulas $\bnabla=(\nabla_z,{R_0}\inv\nabla_\varphi)$ is the
gradient on the unperturbed cylinder. Substituting gives\foot{\mm{This
formula differs from the one in \rNPS\ 
because we no longer need to replace the linear terms using the
ansatz $u=-(u_k)^2+2u_k\cos(kz/R_0)$. Instead we will see how the
terms linear in $u$ determine the constant part of the pressure difference.
This rearrangement of the algebra has no effect on the answers}\kk{,
but the presentation here emphasizes that global volume conservation
is not needed and hence the detailed structure of the terminal blobs
is unimportant.}}
\eqn\eFthree{F[u,\hat\chi]={\kappa\over2R_0^2}\int 2\pi R_0\dd z
\left[{\rm const.}+ (2\sigma-1)u+
\CQ(-{R_0}^2\nabla^2)u^2+\lfr\beta4 \hat\chi^2+\beta(1-k^2)\hat\chi u\right]
\quad,
}
where $\beta\equiv K_0d^2/\kappa$ and
\eqn\eP{\CQ(x)\equiv1+(\sigma-\half)x+x^2+\beta(1-x)^2\quad.}
This is our final formula for the elastic energy. As we
increase the dimensionless tension $\sigma$ (eqn.~\edfsig) from zero,
eventually at some $\sigma\crit$ this quadratic
form \rr{acquires} a negative eigenvalue and the system becomes unstable. (In
Sect.~\Svsm\ we dropped all but the $\sigma$ terms.)

We note in passing that for non-axisymmetric shapes,
$u(z,\varphi,t)=u_{km}\cdot$ $\cos(kz/R_0+m\varphi)$, for $m>1$ the
polynomial \eP\ remains stable until larger values of $\sigma$ than for the
peristaltic $m=0$ modes.

Having found the energy cost $F$
for small distortions in shape and
relative density difference \eFthree--\eP, we now need the dynamical equations for
the two monolayers, the 2d fluids of our four-fluid model.  These equations are the
equations
of tangential force balance.  Each monolayer feels forces due to
inhomogeneities in
density, the traction $T^{\pm}_{\rho z}$ of the 3d water, 2d viscous forces,
and the friction
between the two leaves of the bilayer \SeLa:
\eqn\efb{-\pa_z\fpd {F} {\chi^\pm} \pm T^{\pm}_{\rho z}+
\mu\pa^2_z\vt^\pm\mp b (\vt^+-\vt^-)=0\quad.}
Here $\mu$ is the 2d lipid viscosity; we have written  the frictional force per
unit
area as $b(\dvt)$ where $b$ is a constant and
$\narrowtilde v^\pm$ are the tangential layer velocities
\Evansb\EvYe\SeLa.
Although the change in $\bar\chi$ may be neglected just as in Sect.~\SSbilayer, now
we must keep elastic terms corresponding to changes in $\hat\chi$, since these
are
in principle of the same order as the friction term.
The second and third terms of \efb\ are negligible compared
to the first and the last.  Indeed, since $\vt^\pm\sim\dot u /k$, and
$\dvt\sim {d\over R_0} \dot u /k$, the last
three terms are of magnitude $\eta \dot u$, ${\mu\over R_0}\dot u$, $b
d \dot u$ respectively, so that on micron scales the last term
dominates (see Appendix A).

To get an equation for $\pa_t\hat\chi$, we use the difference in lipid
conservation
laws $\pa_z(\vt^+-\vt^-)=-{d\over R_0} \pa_t\hat\chi$ to write the difference
of the
dominant terms of \efb\ as
$$ \pa^2_z\left(\fpd F {\chi^+} - \fpd F {\chi^-}\right)
= -2 b \pa_z(\vt^+-\vt^-)\quad,$$
or
\eqn\erdd{\pa_t \hat \chi = - {k^2\over {b d^2}}\fpd F {\hat\chi} \quad.}

Next we need the equation of normal force balance analogous to
\fenFB. The Laplace force 
law, eqn.~\epressa, needs to be modified to account for bending
stiffness. To find the right expression we have only to recall the
origin of the Laplace law itself: the pressure jump is $\dl F/\dl V$,
the change in the elastic free energy due to a small change in shape
which changes the volume by $\dl V$. To get \epressa\ we used 
$\fpd FV={\dl F/\dl u\over\dl V/\dl u}$ with 
$\fpd Fu = R_0\Sigma(1-{R_0}^2\nabla^2u)+\cdots)$
and
$\fpd FV = {R_0}^2(1+u+\cdots)$. Now we simply use the full elastic
energy \eFthree--\eP.

We may neglect the difference in monolayer velocities for the purposes of
computing the hydrodynamic factor $\Lambda (k)$. Once again, this
difference is suppressed by
$d/R_0$ relative to the central flow velocity and therefore does not
significantly
modify the energy dissipated in the 3d fluids.
Thus we again have $\dot u$ given by $-\Lambda(k)/\eta$ times the
nonconstant part of the pressure jump, just as in \fetotstrs.
Combining with \erdd, our linearized hydrodynamic equations are
\eqn\eyuck{\pd{}t\pmatrix{ u\cr\hat{\chi}\cr}=
-{\kappa\over R_0^3\eta}\pmatrix{\Lambda(k)&\cr&\epsilon k^2\cr}
\pmatrix{\CQ(k^2)+\half-\sigma& \half\beta(1-k^2)\cr
         \half\beta(1-k^2) & \beta/4 \cr}
\pmatrix {u\cr\hat{\chi}\cr}
\quad,}
\rr{where $\epsilon\equiv R_0\eta/bd^2$ is a measure of the importance of
interlayer friction}.
The desired \mm{dispersion relation is then given by the positive
eigenvalue, if any, of this
linear problem, \ie\ $\omega\sim(-180\,{\rm sec}\inv)\cdot\la$, where}
we took a typical $R_0=0.7$\micron\  and $\lambda$ solves $\lambda^2-
\lambda(g_1+y)+g_0y=0$. Here $g_0=\Lambda(k)[\lfr32 -\sigma+
k^2(\sigma-\half)+k^4]$ is the growth rate with zero friction\rr{,} 
$g_1=g_0+\Lambda(k)\beta(1-k^2)^2$ is the growth rate at infinite
friction\rr{, and $y\equiv \beta\epsilon k^2/4$}. 

\xdef\ftwenty{Fig.~\the\pnfigno}\global\advance\pnfigno by1
\ifigure\ftwenty{Growth rate versus wavenumber in the full linear
model for various values of the dimensionless tension $\sigma$. A
typical experimental value is $\sigma\sim20$.}{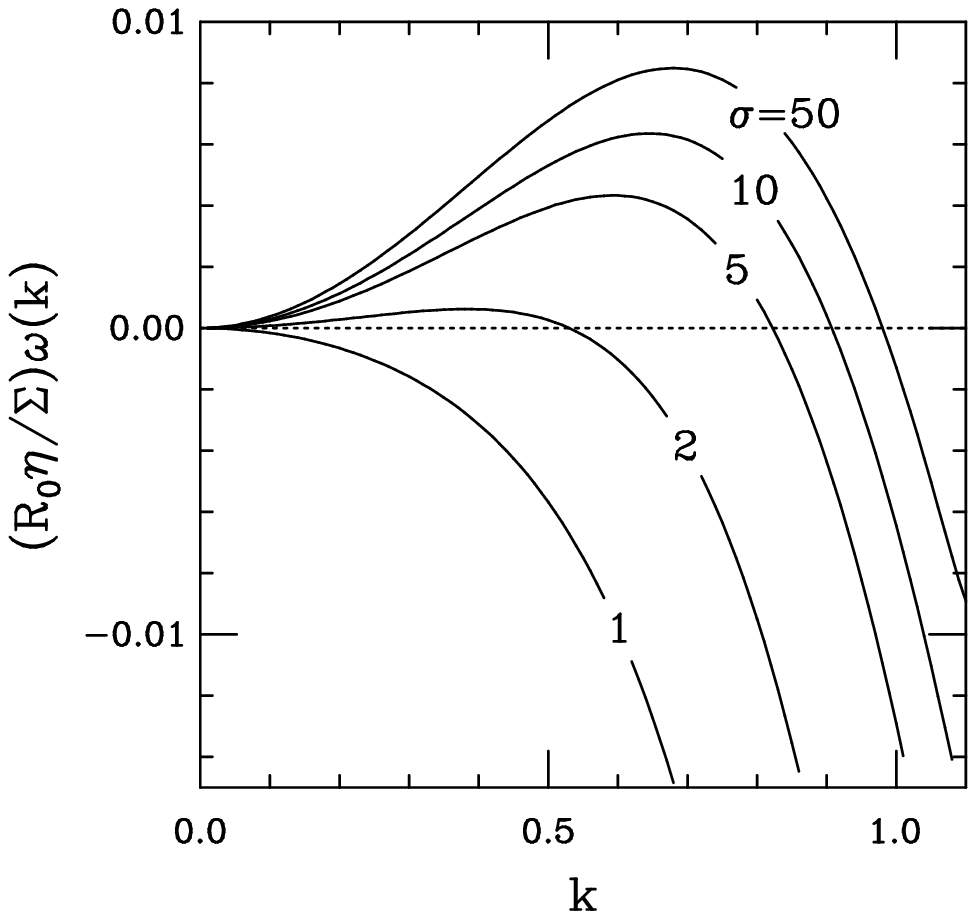}{4.5}

\ftwenty\ shows the growth rate with $\epsilon = .5$, 
$\beta=3.5$, and various values of tension. 
Clearly there is little
effect of stiffness \rr{on the curve $\omega(k)$} when the tension 
\rr{is as large as} its experimental value
$\sigma\sim 20$ \rr{(compare with \fdisp)}. For this value we read off
a fastest growing mode with $k\max=.68$ for
$\sigma=20$,  but as we argued in Sect.~\Sfront\ $k\max$ is {\it
not} the expected wavenumber. Instead
inserting the above formulas for the dispersion $\omega(k)$ into the
equations for the 
MSC \ewvsa--\ewvsb\ gives $k_0=.80$, \rr{a decay length of $4.0R_0$,}
and $v_f=.06 {\Sigma\over\eta}\sim100{\rm 
\mu m}/$sec.
The perfect agreement of $k_0$ with Fig.~1 of \MoBZ\ is fortuitous,
but certainly
the MSC gives reasonable values for the pattern wavenumber and front
velocity, and as promised these values are not very different from
those found in
Sect.~\SSbessel\ since at large tension both the stiffness and the
layer friction are unimportant (see eqn.~\ehierarchy).

\tnewsec\Snon{Model for the Nonlinear Regime} 
We turn finally to a model for the development of
pearls beyond the linear instability.  
A model based on the exact hydrodynamics coupled to the 
elastic forces presents a rather formidable 
computational challenge.  It is therefore of interest to define
a simplified model which nevertheless maintains certain central
features of the full problem.  Chief among these are (i) 
conservation of fluid volume within the vesicle, (ii)
the monotonic decrease in energy associated with low Reynolds number flow,
and (iii) the fully nonlinear structure of
the elastic energy needed to arrive at a true minimum of the energy functional.

The simplest model consistent with these constraints utilizes 
the lubrication approximation, thus yielding a {\it local}
evolution equation for the vesicle radius $r(z,t)$.  
One goal of this investigation is to determine whether propagating fronts of
constant velocity are indeed possible within such a model.  Their
existence would be strongly suggestive that such fronts are supported by
the full hydrodynamics.  A second goal is to test the validity of the marginal
stability criterion.  While the assumptions of lubrication theory may
not be quantitatively correct in comparison with experiment, the issue at hand is
one of internal consistency; within the assumptions of lubrication theory,
how accurate is the MSC?

In the model, the diffusion of tension is ignored, so that
$\Sigma$ is assumed to be uniform along the vesicle.  The bilayer
bending stiffness is included to stabilize pearls against
pinching off, but the small effects of
membrane compressibility and bilayer friction are omitted.
Rather than introducing the optical trap explicitly,
we consider an initial condition of uniform tension with a localized
shape perturbation.  We justified this approach at length in Sect.~3.

We first need some formulas for the elastic force beyond linear
approximation. The  pressure difference across 
the membrane is the 
change in energy with respect to volume, so now our generalization of 
\epressa\ becomes
\eqn\efuncder{\Delta p= {\dl F/\dl r\over\dl V/\dl r}=
{\delta F\over \delta V}= 
2\left\{-\Sigma H +\kappa\left(\hnabla^2 H +2H^3-2HK\right)\right\}\quad,}
where 
$\hnabla^2$ is the
covariant laplacian on the curved surface, and $K$ is the
Gauss curvature.
Axially symmetric shapes described by a radius $r(z)$ (\fcyl) have mean  
and Gaussian curvatures given by
$$\eqalign{H&={1\over2}\left({r_{zz}\over {(1+r_z^2)^{3/2}}}-{1\over {r(1+r_z^2)^{1/2}}}\right)\cr
K&= -{r_{zz}\over {r(1+r_z^2)^2}}~.}$$
Likewise, for these shapes  the determinant $g$ of the
metric tensor and the covariant laplacian are
$$\eqalign{g&=  r\sqrt{1+r_z^2} \cr
\hnabla^2&= {1\over r \sqrt{1+r_z^2}}{\partial\over \partial z} 
{r\over \sqrt{1+r_z^2}}{\partial\over \partial z}~.}$$
For uniform tubular states of radius $R_0$ we see from \efuncder\ that the balance 
of forces yields an equilibrium radius $R_0=(\kappa/2\Sigma)^{1/2}$, as discussed
earlier in the context of the thin ``tethers" between pearls.

Recalling the derivation of the lubrication theory results in Sect.~4,
in particular Eq. \econswater\ , we know that the equation of motion for
the radius $r(z,t)$ should have the form of a local conservation law,
\eqn\eluba{{\pa\over {\pa t}}\pi r^2= -{\pa J\over{\pa z}}\quad,}
where $J$ is the axial current.  For this we appeal to our previous 
(linearized) result on lubrication theory that relates $J$ to the gradient 
of pressure, 
\eqn\elubb{J= -{\pi \over 4\eta}r^4 {\partial P\over \partial z}~.}
Recall that the factor of $4$ in the denominator differs from the usual
Poiseuille result due to the imposition here of lipid conservation.
The pressure is just that determined by the energy functional, eqn.~\efuncder.
Combining Eqs. \eluba, \elubb, and \efuncder, and rescaling both $r$ and
$z$ with the unperturbed radius $R_0$ and introducing the rescaled time
$\tau=(\eta R_0^3/4\kappa)t$ we obtain the partial
differential equation
\eqn\enifty{r{\pa r\over {\pa \tau}}= - {\pa \over{\pa z}}
\left[r^4{\pa \over{\pa z}}\left(
-\sigma H +\hnabla^2 H +2H^3-2HK\right)\right]\quad ,}
where $\sigma$ is again the rescaled tension \esigmaest.

It is easily demonstrated that this is a gradient flow, for the time derivative of
the functional $F$ is
$$\eqalign{{\dd F\over \dd\tau}&= \int\dd z {\delta F\over \delta r}{\pa r\over {\pa \tau}}
= \int\dd z {\delta F\over \delta r}{1\over r}\partial_z\left(r^4\partial_z
{1\over r}{\delta F\over \delta r}\right)\cr
&= -\int\dd z\, r^4 \pa_z\left({1\over r}{\delta F\over \delta r}\right)^2 \le 0~,}$$
where the last line follows by an integration by parts.  Thus, provided $r>0$
(i.e. the interface does not pinch off), $F$ is driven strictly downhill.
Furthermore, when $F$ is constant in time the functional derivative
$\delta F/\delta r=0$ and the system is at an energetic extremum.
Since the model \enifty\ also contains the proper linear stability
result of the lubrication approximation, it provides a dynamics that
interpolates between the basic Rayleigh-like instability and the stationary final
states, while obeying the relevant hydrodynamic conservation laws.

The flux form of the equation of motion and the relation between
pressure gradients and velocity in \enifty\ are features found as well in
models introduced recently for interface motion leading to topology
transitions and singularities in viscous flows \rpinching.  
In all of these systems, despite the simplicity of the physical ingredients, 
the equation of motion is highly nonlinear and of very high degree 
(sixth order in $z$-derivatives in the present case).
These features make numerical studies quite delicate, but with care a stable
and accurate algorithm may be developed.  Here we report preliminary results
of simulations illustrating that propagating fronts of peristalsis are
supported by the model.  

\tfig\fpropa\ shows the evolution of a cylindrical vesicle with $\sigma=10$
perturbed initially with a localized distortion.  The computational domain
has a length $16\times 2\pi$, and periodic boundary conditions are
imposed by use of a pseudospectral algorithm.  The figure clearly shows
that following an initial ``induction" period there is a propagating
peristaltic pattern moving symmetrically outward from the initial
perturbation.
{A nonlinear least-squares fit of the leading edge of the front
at successive times to the product of a decaying exponential and a
cosine function shows that the position of the front increases linearly 
with time over the duration of the simulation, during
which it has moved about six pearl lengths.  
This constant velocity is in accord with our physical arguments of
Sects. 3--\Svsm . 
We find the dimensionless wavenumber of the pattern in the leading edge of
the front to be $k_0\approx 0.78$, and the  decay length to be $\approx 5.4$.  
Applying the techniques of the earlier sections to the model of this
section, one finds that the MSC predicts a wavenumber $k_0=0.77$;
the decay length of $4.3$ is somewhat smaller than the value
observed in the numerical solution,
but is still on the order of a pearl size.  Thus the front is ``sharp." 
Finally, in the rescaled units of \enifty, the front velocity at the
value $\sigma=10$  is found to be $v_f\approx 13$, close to but
somewhat smaller than the marginal stability criterion value of $17$. 
Work is in progress to understand the origins of these
discrepancies.} 

\ifx\answ\bigans
\ifigure\fpropa{Numerical solution of the lubrication equation \enifty\ with
$\sigma=10$, showing the formation of pearls.  Time increases from
bottom to  
top in increments of $\Delta \tau=1.0$.}{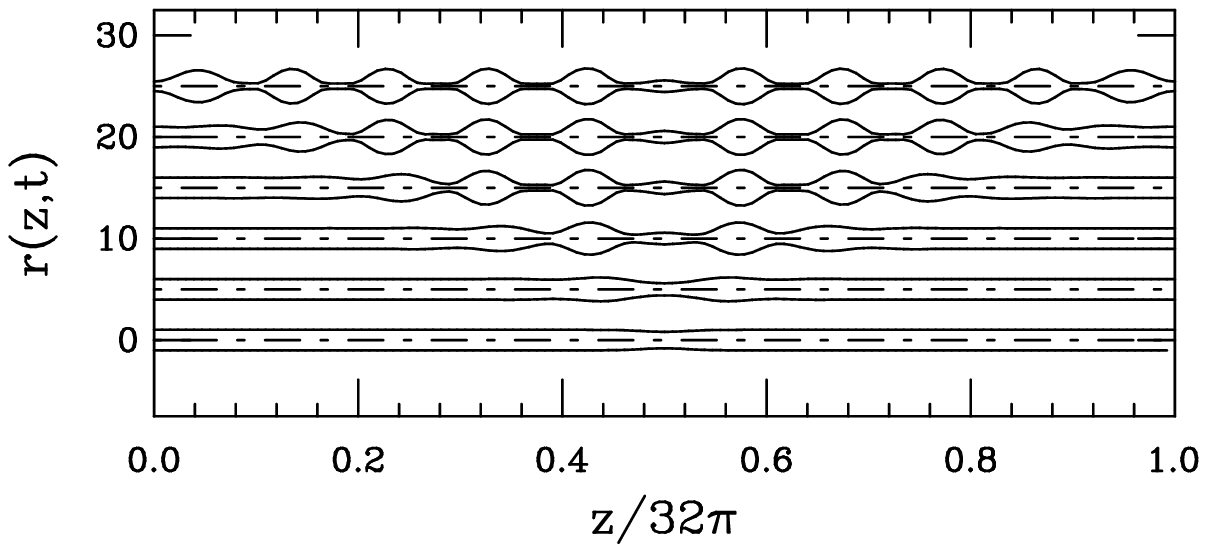}{3.2}
\else
\ifigure\fpropa{Numerical solution of the lubrication equation \enifty\ with
$\sigma=10$, showing the formation of pearls.  Time increases from
bottom to  
top in increments of $\Delta \tau=1.0$.}{propa.eps}{2.5}
\fi

In addition to these quantitative aspects of front propagation, we see
that the state left behind the front indeed has the appearance of a string of
pearls, with very narrow tethers between somewhat prolate ellipsoids.  These are
quite similar to ones seen in the experiments of Bar-Ziv and Moses.
It is also interesting to note that a slight shift between the
selected wavelength near the leading edge of the front and that
found well behind, where the pearls are fully formed.  This difference
is in qualitative accord with what one finds in other more familiar 
examples of propagating pattern selection \DeeL .

\newsec{Conclusion}
Laser
tweezers are an important new experimental tool for understanding
membrane dynamics; they furnish a
precisely controllable, local, physical intervention. \kk{In this
paper and \rNPS\ we have proposed a 
simple model for the effective laser-membrane interaction and
used it to explain much of the phenomenology of the pearling
instability discovered in this context by Bar-Ziv and Moses.}

Various aspects of the pearling phenomenon remain to be explained.
For example, 
we have not attempted to study the {\it migration} of the pearls which
develop after 
prolonged tweezing. 
What we did show was how a cylindrical vesicle under tension is unstable
to states with a periodic modulation in diameter. 
We refined our previous prediction for the initial wavelength
of this perturbation using the marginal
stability criterion and predicted a
selected wavenumber in good agreement with the experiments of
ref.~\MoBZ. We also explained the observed front behavior and 
predicted a front velocity in qualitative agreement with the experiments.
Along the way we showed how to incorporate the bilayer membrane
structure into the dynamics.  Although
this structure did not significantly affect our results, it may matter
in dynamics problems at lower tension and on shorter length
scales. Finally, our numerical calculations have demonstrated that
the lubrication approximation is rich enough to capture the essential
physics of the pearling phenomena, such as the existence of an initial
uniform front propagation velocity and the shape of the fully developed 
pearls. These calculations were in reasonable agreement with the 
predictions of the MSC, thus bolstering our faith in this approach and
suggesting that the MSC can be reasonably applied to other fluid dynamics
problems.  One particularly relevant problem is the propagation of
the ordinary Rayleigh instability along a column of fluid, as seen
in recent experiments on droplet fission \ShiBrenNagl .

We are not aware of any naturally-occurring biological process to
which the  pearling instability literally applies. Our goal was rather
to create a simple theory of membrane dynamics and test its many
underlying assumptions by applying it to a well-controlled dynamical
shape transformation. We expect that the approach developed
in this paper can provide a starting point for understanding other
phenomena of more direct biological interest, for instance the
dynamics of shape instabilities during adhesion 
and budding
.


{\frenchspacing
\ifx\prlmode\testp\else\vskip1truein \leftline{\bf Acknowledgements}
\noindent\fi
We would like to thank
R. Bruinsma,
F. David,
B. Fourcade,
M. Goulian,
R. Granek,
J. Krug,
R. Lipowsky,
T. Lubensky,
X. Michalet,
S. Milner,
D. Nelson,
Z. Olami,
V.A. Parsegian,
S. Safran,
and S. Svetina for innumerable discussions.
 We especially thank R.
Bar-Ziv and E. Moses for constant help and in particular for
suggesting the applicability of the MSC.
PN\ also thanks CEA Saclay, the Weizmann Institute, and the Institute
for Advanced Study, and PN\ and US thank ITP Santa Barbara for
their hospitality and partial support, while  some of this work was done.
REG acknowledges support from NSF PFF Grant DMR93-50227 and the
Alfred P. Sloan Foundation. PN acknowledges NSF
grants PHY88-57200
and DMR95-07366,
and the Donors of the Petroleum Research Fund,
administered by the American Chemical Society, for the partial support
of this research.}

\tappendix\SAconsts{A}{A Compendium of Constants}
{\obeylines\parindent0pt
\noindent Here we list the values of various physical constants for DMPC %
bilayers. We take:
the bilayer bending stiffness to be $\kappa=0.6\eem{12}$\erg\ %
\rEvNe,
the bilayer thickness to be $2D=40$\AA\ \rAskUdo,
the distance from bilayer midplane to each monolayer neutral surface %
to be $d\sim D/3=12$\AA\ \WSSZ\MSWD,
the bare bilayer compression modulus to be $K_0=144\,$\erg\cmmt\ \rEvNe,
the interlayer friction constant $b$ to be $10^8$\erg sec%
\thinspace cm$^{-4}$ \Evansb,
the viscosity of water $\eta$ to be $1\eem2$\erg sec\thinspace cm$^{-3}$,
and the 2d viscosity of the lipid layer to be $\mu=10^{-6}$\erg sec$\,$cm$^{-2}$%
\rMeSaEv.

In addition for concreteness we sometimes take:
the typical tubule radius $R_0=0.7$\micron (ref. \MoBZ, fig 1),
the typical laser power 50mW \MoBZ,
and the  temperature to be $k_B T=4.2\eem{14}$\erg.

Thus we get the derived quantities:
$k_B T/4\pi\kappa=0.006$,
$\Sigma\crit\equiv\kappa/{R_0}^2=1.2\eem4$\erg\cmmt,
$\epsilon\equiv R_0\eta/bd^2\sim0.5$,
$\beta\equiv K_0d^2/\kappa\sim3.5$.

Also, we estimated the induced tension $\Sigma\sim2\eem3$\erg\cmmt\ in %
Sect.~\FFlaser, so
$\Sigma/\eta\sim2\ee3$\micron/sec and the dimensionless tension
$\sigma\equiv\Sigma{R_0}^2/\kappa\sim20$.
The values of $\Sigma,b,d$ are not known very accurately, but fortunately our %
formula for $k_0$ is not very sensitive to them.

}

\tappendix\SAdelau{B}{The Delaunay Surfaces}
We argued in the text against the introduction of
a spontaneous curvature term in the elastic free
energy. In the presence of such a term we may add and subtract a
constant to recast the energy in the form
$$F[{\rm shape}]=2\kappa\int\Bigl[\bigl(H-H_0\bigr)^2+\sigma'\Bigr]\quad.$$
Large thermal fluctuations then imply that initially $\sigma'=0$, and
the free energy then simply prefers conformations with {constant
mean curvature} everywhere equal to $H_0$.

Remarkably there is a {\it one-parameter family} of such shapes with
axial symmetry. These are the ``unduloid'' or ``Delaunay'' surfaces.
They obey a nonlinear equation expressing $H=H_0$. Deuling and
Helfrich used this equation in ref.~\DeHe. There is however a very
simple {\it geometrical} characterization of these surfaces~\Bouasse
:
\itemitem{}{\sl Draw a line $\ell$ in the plane. Construct any ellipse,
tangent at one point to $\ell$. Let the ellipse roll without slipping
along $\ell$, and follow the path of one focus to get a curve $C$. The
figure of revolution obtained by rotating $C$ about the axis $\ell$ is
a surface of constant mean curvature $1/2r_+$, where $r_+$ is the
larger semimajor axis of the ellipse.
}

Let us see what this says for nearly-cylindrical
shapes. Such shapes are generated from nearly-circular ellipses, with
foci near the center. The
radius $R_0$ of the near-cylinder is thus equal to that of the near-circle.
Rolling the near-circle along the axis of course gives a periodic
shape with period  $\lambda=2\pi R_0$. In our language, this is a small
perturbation with wavenumber $k=1$, larger than the initial wavenumber seen
experimentally.

\optional{To prove the theorem, draw a tangent to the curve in the plane at a
point $B$, and extend it until it intersects $\ell$ at an angle
$\theta$, at $F$. See \tfig\funduloid. Also drop a perpendicular $BE$ from
the tangent to $\ell$, and another $DB$ from $\ell$ to the tangent.
Let $y=|EB|$.

\ifigure\funduloid{Construction of the unduloid surface.}{unduloid.eps}{2.5}

One principal curvature to the surface of revolution lies in the plane
a distance $1/c_1$ from $B$ along $BD$. A small excursion d$s$ in
arclength from $B$ along the curve then changes the tangent angle to
$\theta+\dd\theta$.  To lowest order in d$s$ we have $\dd\theta=-c_1\dd
s$ since the center of the osculating circle only moves at second
order. The other principal curvature direction comes out of the plane;
the osculating circle is formed by point $B$ when we rotate the segment
$BD$ about the $\ell$ axis. So the other principal curvature is
$c_2= 1/|BD|$, and the constant-curvature equation is
$2H=-{\dd\theta\over\dd s}+{\cos\theta\over y}=$const.

We can simplify by noting that d$y=\dd s\sin\theta$, so
$2H\lfr d{dy}(y\cos\theta)=y$ or $\cos\theta={y\over 4H}+{a\over y}$ where $a$
is a constant of integration. Letting $n=y/\cos\theta$ be the length
of $BD$, we get
\eqn\edelaunay{y^2(\lfr1n-\lfr1{4H})=a\quad.}

As in the statement of the theorem let us construct an ellipse with
semimajor axes $r_+=(2H)\inv$, $r_-=\sqrt{4aH}$, tangent to $\ell$ at
$D$ and with one focus lying on our curve at $B$, the other at $B'$.
Thus the two foci are separated by a distance $2\sqrt{r_+^2-r_-^2}
=4\sqrt{H^2-aH}$.
Letting the length of $B'D$ be $n'$, we have $n+n'=2r_+$ by the
definition of an ellipse.

We also find that the angle $\angle BDB'=2\theta$. To see this, draw
an ellipse with foci $B,B'$ and a tangent line at $D$. Let the tangent
make angles $\alpha,\beta$ with the segments $BD$ and $B'D$
respectively. Moving along the tangent a distance d$s$, distance $|BD|$
increases by d$s\cos\alpha$, while $|B'D|$ decreases by d$s\cos\beta$,
so to preserve a constant sum to first order we must have
$\alpha=\beta$. In our
figure $\alpha$ is the complement of $\theta$, and since the angle
$\angle BDB'$ is the sum of the complements of $\alpha$ and $\beta$
it must equal $2\theta$.

Using $n+n'=2r_+$, $\cos\theta=y/n$, and the distance between foci
given above, we then at once find $a={y^2\over n}-{y^2\over 4H}$,
which is the desired \edelaunay.
}

\tappendix\SAmodulus{C}{Thermal Fluctuations}

\optional{Our linearized elastic energy \eFthree\ contains the physical
constant $\beta\equiv K_0d^2/\kappa$, which expresses the resistance
to {\it differential} stretching of the layers in terms of the resistance
to bending. We argued in Sect.~3 that for the linear theory the
resistance to {\it overall} stretching would not enter the linear
theory, as long as it was large enough. We now estimate this modulus,
partly to justify our formula \eFthree\ and partly for use
in \MSCsect.}

In subsection \SSinprop\ we quoted an estimate for the effective
stretching modulus of the bilayer. 

Generally the effects of thermal fluctuations on the conformations of
stiff membranes are quite small unless one carefully adjusts to the
threshold of a shape transformation. One simply replaces the elastic
constants by effective values corrected by terms of relative order
$k_BT/4\pi\kappa$, about half a percent for DMPC.\foot{Something similar
happens with {\it dynamic} couplings \TCLWC.} We will neglect such
effects. One notable exception, however, is the susceptibility
$K\inv$, whose ``bare'' value is so small that thermal corrections can
be significant. At modest values of tension, the membrane effectively
becomes elastic, because significant area can hide in invisible
short-scale wrinkles excited by thermal agitation \HeSe\EvRa. (The tiny
value reported for the bare $K_0\inv$ in \EvRa\ is attained only at
extremely large values of applied tension, where these wrinkles have
been extinguished.)

Following Helfrich and Servuss, let us estimate the magnitude of this
effect in our case. We consider a membrane thermally fluctuating about
a cylindrical shape, and evaluate the average true area
$A_0\equiv\langle\int\dd S\rangle$ in terms of the ``projected'' or apparent
area $2\pi R_0L$. We would like to know how $A_0$ changes as we adjust
the tension, at fixed projected area; this tells us how much area we
can pull out of our membrane at fixed apparent shape, and hence the
effective modulus. Following \HeSe, the relative area extension as we
change the dimensionless tension $\sigma$ is
$$\dl A/A\sim {k_B T\over{8\pi^2}}\int_{\sqrt{\sigma_0}}^{\sqrt\sigma}{\dd k\
k^2\over\kappa k^4}$$
In our case $\sigma\sim20$ (Appendix~\SAconsts). The lower bound of the integral should be
either the initial tension or unity, whichever is larger, to account
for the cylindrical geometry; since we take the initial tension to
vanish and $\sigma\sim20$ we get $\dl A/A\sim (k_BT/8\pi\kappa)\log 20\sim
$1\%. This
estimate is not very sensitive to the actual value of $\sigma$.
Then the effective modulus is given by the constitutive relation
$\Sigma=K\eff (\dl A/A)$, or
$K\eff\sim100\cdot\Sigma\sim2\eem1$\erg\cmmt, so that
the hierarchy \ehierarchy\ we assumed earlier is still preserved.

\tappendix\SAlambda{D}{Computation of the Dynamical Factor $\Lambda(k)$}

Eq. \eNSstr\ is solved by separating
variables as $\psi=\Psi(\rho)\exp(\omega t + i k z/R_0).$  $\Psi$ 
satisfies
$(\pa_\rho^2-\rho^{-1}\pa_\rho-(k/R_0)^2)^2\Psi=0$, which has solutions of the form
\eqn\eMBssls{\Psi(\rho)=A_1\rho{\rm I}_1\left(k\rho/R_0\right)+
B_1\rho {\rm K}_1\left({k\rho/R_0}\right)
+A_2\rho^2 {\rm I}_0\left({k\rho/ R_0}\right)+B_2\rho^2 
{\rm K}_0\left({k\rho/R_0}\right)\quad,  }
 where ${\rm I}_\nu(x), {\rm K}_\nu(x)$ are modified Bessel functions 
\AbSt.
Demanding that $\Psi(\rho)$ be well-behaved at $\rho=0$ and $\rho=\infty$
forces
$B_1=B_2=0$ in the interior fluid and $A_1=A_2=0$ in the exterior fluid.
The no-slip boundary conditions, together with the lipid conservation law
\elipconsb,
give us four equations for the four unknowns $A_1, A_2, B_1, B_2$:
\eqn\enoslip{v_\rho^+(\rho=R_0)=R_0\dot u\ ,\qquad v_\rho^-(\rho=R_0)=R_0\dot
u\ ,}
\eqn\elpcon{v_z^+(\rho=R_0)=\vt={R_0\over i k}\dot u\ ,\qquad
v_z^-(\rho=R_0)=\vt={R_0\over i k}\dot u\quad, }
 where $+$ ($-$) denotes the exterior (interior) fluid.  Note that in \elpcon\
we have assumed that the lipid velocity $\vt$ has no constant part, in accord
with the arguments of Sect.~\Fphys\ that after the tension has spread the 2d
pressure (tension) is practically uniform and we can take any net flow of lipid
to be small.
Also, we have taken the lipid to be
incompressible since the applied tension is much smaller than the membrane
compressibility (eqn.~\ehierarchy).
Plugging \eMBssls\ into \enoslip--\elpcon, and using the relations
\eqn\eBI{{\rm I}'_0(x)={\rm I}_1(x)\ ,\qquad {\rm I}'_1(x)={\rm I}_0(x)-{1\over
x}{\rm I}_1(x)\quad,}
 \eqn\eBK{{\rm K}'_0(x)=-{\rm K}_1(x)\ ,\qquad {\rm K}'_1(x)=
\rr{-}{\rm K}_0(x)-{1\over x}{\rm K}_1(x)\quad,} we find
\eqnn\efoureqns
$$
\eqalignno{i {k\over R_0} (A_1{\rm I}_1(k)+A_2 R_0{\rm I}_0(k))=&R_0\dot u \cr
i {k\over R_0} (B_1 {\rm K}_1(k)+B_2 R_0 {\rm K}_0(k))=&R_0\dot u\cr
A_1 {k\over R_0} {\rm I}_0(k)+A_2(2 {\rm I}_0(k)+k{\rm I}_1(k))=&{R_0\over i k}\dot u\cr
B_1 {k\over R_0} {\rm K}_0(k)+B_2(k{\rm K}_1(k)-2 {\rm K}_0(k))=&{R_0\over i k}\dot
u\quad,&\efoureqns}
$$
{\it i.e.} the $A$'s and the $B$'s are proportional to $\dot u$, with
proportionality
constants which are rational functions of the modified Bessel functions.

To relate $\dot u$ to $u$ and thus determine the growth rate $\dot u/u$, we
examine the normal force balance equation at the membrane:
\eqn\enFB{T^+_{\rho\rho}-T^-_{\rho\rho}-\Delta p\lap=0\quad,}
where  $T_{ij}$ is the 3d stress tensor of water and $-\Delta p\lap$
is the pressure jump due to the surface 
tension, given by \epressa.  The $z$ component of the Navier-Stokes
equation relates the pressure to the velocity,
$p^{\pm}={\eta R_0\over{i k}}\nabla^2 v^{\pm}_z+p^{\pm}_0$. The NS
equation does not fix the constants $p^\pm_0$; we must choose $p^+_0=0$
and $p^-_0=\Sigma/R_0$.
Using
$$\left(\rho\inv\pa_\rho \rho\pa_\rho+\bigl(\lfr k{R_0}\bigr)^2\right)f=0$$
for $f(\rho)={\rm I}_0({k \rho\over R_0})$, 
${\rm K}_0({k\rho\over R_0})$ \AbSt, we find
at the membrane\foot{\kk{We have used $T_{ij}=-p\delta_{ij}+\eta(
\pa_iv_j+\pa_jv_i)$ for incompressible fluids \ref\LLfluid{L. Landau
and E. Lifsitz, {\sl Fluid mechanics} (Pergamon, 1959). }.}}
\eqna\etotstrs
$$
\eqalignno{T^+_{\rho\rho}-T^-_{\rho\rho}=&-2i\eta {k\over R_0} 
(B_1{k\over R_0} ({\rm K}_0(k)+{1\over
k}
{\rm K}_1(k))+B_2k{\rm K}_1(k))\cr
&-2i\eta{k\over R_0} (A_1{k\over R_0}({\rm I}_0(k)-{1\over k}
{\rm I}_1(k))+A_2k{\rm I}_1(k))+\Sigma/R_0&\etotstrs a\cr
\narrowequiv&-\eta\Lambda(k)\inv\dot u + \Sigma/R_0\quad.&\etotstrs b\cr}
$$
Eqn.~\etotstrs b defines what we call the ``dynamical factor''
$\Lambda(k)$ in Eq. \eGR.
Taking the values for the $A$'s and $B$'s 
determined by \efoureqns, we find 
\eqn\ekinfac{\Lambda (k)=-{1\over 2k}{
[k(K_0^2-K_1^2)+2K_0K_1][k(I_0^2-I_1^2)-2I_0I_1]\over
3I_0K_0/k  - K_1I_1/k+ k(I_1^2K_0^2-I_0^2K_1^2)}\quad.}
In eqn.~\ekinfac\ all Bessel functions are to be evaluated at $k$.

\tappendix\SAade{E}{Membrane Elasticity}
Here we briefly derive eqn.~\eADE. Following previous authors (\eg\
\MSWDetc\Mich) we visualize each monolayer as a fluid of compressible
cylinders, each of length $D$. One end (the ``chain'') touches the
bilayer midplane and resists compression and extension with a modulus $K_c$ and
preferred area/molecule at the end of $a_{c0}$.\optional{\foot{Of course the
elasticity is really distributed along the chain, but in the end our
discrete model gives the same answers as a more elaborate one with
distributed modulus.}}{} The other end (the
``head'') is a normal distance $\pm D$ from the midplane with modulus
$K_h$ and preferred area $a_{h0}$. All of the parameters introduced
are the same for each monolayer, since the two layers and their
respective solvents are identical.
We thus consider a monolayer
elastic energy per molecule of
\eqn\eFmono{f^\pm[{\rm shape,} a_h^\pm,a_c^\pm]
=\half K_ha_{0h}\left({a_h^\pm\over a_{0h}}-1\right)^2
+\half K_ca_{0c}\left({a_c^\pm\over a_{0c}}-1\right)^2 - \const\quad.
}
Here the constant includes the chemical potential for extracting
additional molecules from any reservoir; it
will of course turn out to be related to the
physical tension.

It's convenient to introduce a pair
of imaginary ``neutral'' surfaces at normal distance $\pm d$ from the
midplane, where $d$ will be specified in a moment, and to measure
the densities $\phi^\pm$  of lipid molecules in the two monolayers
relative to these surfaces. Thus the areas $a^\pm_{h,c}$ above are
$$\eqalign{a_h^\pm&=(\phi^\pm)\inv\bigl[1\mp2H^\pm(D-d)\bigr] \cr
a_c^\pm&=(\phi^\pm)\inv\bigl[1\pm 2H^\pm d\bigr] \quad.\cr}$$
Here $H^\pm$ are the mean curvatures of the two neutral surfaces.
We dropped terms involving Gauss curvature, as well as those of order
$(HD)^2$. Substituting in \eFmono\ we get
\eqn\eFmonob{f^\pm=(\phi_0)\inv\left[
{\kappa\over4}\bigl(2H^\pm\mp c_m\bigr)^2
\mp\kappa c_m\left({\phi^\pm\over\phi_0} -1\right)\cdot2H^\pm
+{ K\over4}\left({\phi^\pm\over\phi_0} -1\right)
\right]-\const'\quad,
}
where $K,\phi_0,\kappa,c_m$, the new constant, and the neutral surface
location $d$ are various combinations of the original parameters
chosen to put \eFmono\ into the form \eFmonob. In particular $d$ is
chosen to set the coefficient of the cross-term in the convenient form
shown.  Since the original
parameters weren't directly observable anyway, we abandon them now in
favor of the new ones. This is actually progress. Note that again the
new parameters are the same for each monolayer.

The monolayer free energy is now $\int\dd S_\pm\,\phi^\pm f^\pm$,
where $\dd S_\pm$ are the areas of the two neutral  surfaces.
Rewriting in terms of the area element $\dd S$ of the midplane we get
$$\eqalign{\dd S_\pm\,\phi^\pm f^\pm&=\dd S(1\mp 2Hd)\cr\times&\left[
{\kappa\phi^\pm\over4\phi_0}\bigl(2H^\pm\bigr)^2
\mp\half\kappa c_m(2H^\pm)
+{\phi^\pm\over\phi_0}{K\over4}\left({\phi^\pm\over\phi_0}-1\right)^2
-\const^{\prime\prime}\cdot\phi^\pm
\right]\quad.\cr}$$
Here $H$ is the mean curvature of the midplane: $H^\pm=H\pm
2H^2d$ plus a Gauss-curvature term we drop.

We will be considering only values of the tension much smaller than $K$
(eqn.~\ehierarchy), so we may use  $\phi_0$ in place of $\phi^\pm$ in
the constant term above. Writing the other $H^\pm$ in terms of $H$ and
finally identifying the constant with the tension $\Sigma$
then yields eqn.~\eADE.

Before switching on the laser the terminal blobs serve as reservoirs,
and we argued in the text that then $\Sigma$ should be taken to be
close to
zero. In equilibrium the densities will then just take their preferred
values, and we recover the Canham-Helfrich model, with $\kappa$ the
bilayer stiffness. In particular, no spontaneous curvature appears, as
argued in \Evansb. After switching on the laser the system is no
longer in equilibrium, and we  need to work out dynamically what
happens.

We note in passing that in a {\it closed} system the constant term
in \eFmono\ will {\it not} be equal for the two layers; rather than
containing a chemical potential for a reservoir, these terms act as
Lagrange multipliers enforcing the constraint of fixed molecule
numbers $N^\pm$ in each layer. Letting $A_0^\pm\equiv a_0N^\pm$ be the
``preferred area'' of each layer, we then recover the
area-difference elasticity model (ref. \MSWD). In particular the
parameter $\alpha_{ADE}$ is essentially the same as our parameter
$\beta$.

\footatend\bigskip\immediate\closeout\rfile\writestoppt
  \baselineskip=22pt\centerline{{\bf References}}\bigskip{\frenchspacing%
  \parindent=20pt\escapechar=` \input refs.tmp\vfill\eject}\nonfrenchspacing
 \vfill\eject\immediate\closeout\ffile{\parindent40pt
 \baselineskip22pt\centerline{{\bf Figure Captions}}\nobreak\medskip
 \escapechar=` \input figs.tmp \vfill\eject
}

\bye